\documentclass[11pt,letterpaper]{article}
\ifdefined\pdfminorversion\pdfminorversion=7\fi
\usepackage[margin=1in]{geometry}
\usepackage[T1]{fontenc}
\usepackage{lmodern}
\usepackage{microtype}
\usepackage{amsmath,amsfonts,amssymb,amsthm}
\usepackage{algorithm,algpseudocode}
\usepackage{graphicx,placeins,xcolor}
\usepackage[authoryear,round]{natbib}

\usepackage[font=small,labelfont=bf]{caption}
\usepackage[hidelinks]{hyperref}
\hypersetup{
  pdftitle={Empirical Bayes mirror testing for persistent widening of achievement gaps in U.S. school districts},
  pdfauthor={Bowen Gang and Yin Xia},
  pdfsubject={},
  pdfkeywords={Achievement gaps, Empirical Bayes, False discovery rate, Mirror statistics, Multiple testing}
}
\newcommand{\rev}[1]{{#1}}
\numberwithin{equation}{section}
\newcommand{\R}{\mathbb R}
\newcommand{\E}{\mathbb E}
\newcommand{\Normal}{\mathcal N}
\newcommand{\FDR}{\operatorname{FDR}}
\newcommand{\lfdr}{\operatorname{lfdr}}
\newcommand{\FDP}{\operatorname{FDP}}
\newcommand{\Prob}{\mathbb P}
\algrenewcommand{\algorithmicrequire}{\textbf{Input:}}
\algrenewcommand{\algorithmicensure}{\textbf{Output:}}
\theoremstyle{plain}
\newtheorem{proposition}{Proposition}[section]
\newtheorem{theorem}{Theorem}[section]
\theoremstyle{definition}
\newtheorem{definition}{Definition}[section]

\title{Empirical Bayes mirror testing for persistent widening of achievement gaps in U.S. school districts}
\author{Bowen Gang\textsuperscript{1}\qquad Yin Xia\textsuperscript{2}\\[0.6em]
  {\small\textsuperscript{1}Department of Mathematical and Statistical Sciences,}\\
  {\small University of Alberta}\\[0.3em]
  {\small\textsuperscript{2}Department of Statistics and Data Science, Fudan University}\\[0.4em]
  {\small\texttt{bgang@ualberta.ca}\qquad\texttt{xiayin@fudan.edu.cn}}}
\date{}

\begin{document}
\maketitle

\begin{abstract}
Unequal recovery from pandemic learning losses raises concerns about
widening achievement gaps between economically disadvantaged students
and their peers. \rev{This study seeks to identify districts whose
mathematics gaps widened in every annual interval from 2022 to 2025.
The analysis uses estimates from the Stanford Education Data Archive
(SEDA) for 5,600 U.S. school districts. Identifying such districts
requires accounting for correlated errors and differing precision in
the estimated annual changes, as well as} \textcolor{black}{multiplicity}.
\textcolor{black}{The true annual gap changes in each district form an
effect vector, and persistent widening means that every component is
positive.}
\textcolor{black}{To address this class of multiple testing problems,
we propose an empirical Bayes mirror procedure.}
\rev{It uses}
an estimated distribution of effect vectors to guide ranking and
covariance-adjusted mirror observations to calibrate selection.
\rev{The procedure achieves finite-sample false discovery rate (FDR) control}
under a Gaussian observation model, \rev{without requiring accurate estimation
of the posterior null probabilities used for ranking}. Simulations demonstrate power
gains from incorporating information about the distribution of
\textcolor{black}{effect vectors}. At a target FDR of 25\%, the SEDA analysis identifies eight
districts in five states with evidence of widening gaps in every annual
\rev{interval.}
\end{abstract}


% BEGIN SECTION 1
\section{Introduction}
\label{sec:introduction}

Growing educational inequality after COVID-19 raises concerns about the
future economic opportunities of children from disadvantaged families.
Pandemic learning losses were larger among children from lower
socioeconomic backgrounds \citep{betthauser2023}, and the average
within-district socioeconomic mathematics gap in the United States increased from
2022 to 2024 \citep{dewey2025}. We analyze the Stanford Education Data
Archive (SEDA) 2025.2 \citep{seda2025}, which provides annual achievement
estimates and standard errors. Our sample includes 5,600 school
districts in 33 states observed from 2022 to 2025. \rev{The central question is which districts}
experienced a widening mathematics gap between economically
disadvantaged students and their peers in every annual interval.
Since academic achievement is linked to college attendance and adult
earnings \citep{chetty2014}, identifying these districts can help locate
communities where \textcolor{black}{the achievement gap between disadvantaged
students and their peers continues to widen}.

\rev{Recovery patterns vary substantially across districts \citep{dewey2026}.
Within a district, a larger gap at the end of the period does not by itself
establish an increase in every year. Assessing such persistence requires
accounting for differences in the precision of annual estimates across
districts and for correlated errors in adjacent estimated changes.
These correlations arise even when annual estimation errors are
independent, because a shared annual estimate enters the two differences
with opposite signs. With thousands of districts under consideration,
multiple comparisons must also be addressed. We therefore seek to
identify persistent increases while controlling the false discovery
rate (FDR\textcolor{black}{; \citealp{benjamini1995}}), the expected proportion of selected districts for which
the claim of persistent widening is false.}

\rev{A standard approach takes the largest one-sided $p$-value across
the annual changes for each district}
\citep{benjamini2008} and applies the Benjamini--Hochberg procedure
\citep{benjamini1995}. Although this approach controls FDR under
independence across districts, its ranking depends only on the weakest
standardized change. \rev{It therefore leaves information beyond the
weakest change unused and does not adapt the ranking to the distribution
of change vectors across districts.}

\rev{\textcolor{black}{The same testing problem arises beyond achievement
gaps whenever the goal is to identify units whose mean vectors have
all positive components.} For example,
replicability analysis may seek genes with positive effects in each of
several independent studies \citep{benjamini2008,heller2014}. In
dose-response screening, one may seek genes whose mean expression
increases at every dose increment for a fixed drug, cell type, and
exposure time. Large drug-perturbation resources such as LINCS L1000
provide expression profiles for investigating such questions
\citep{subramanian2017}. When the dose-specific means are estimated
independently, their adjacent differences have the same covariance
structure as the annual changes in SEDA. These problems share a union
null hypothesis that allows at least one nonpositive effect, with the
remaining effects unrestricted.}

\textcolor{black}{Empirical Bayes methods are widely used in large-scale
multiple testing. A key idea is to}
\textcolor{black}{use information across units to}
\textcolor{black}{rank hypotheses by the local false
discovery rate (lfdr), the posterior probability that the null hypothesis
is true \citep{efron2001,sun2007}. Extensions incorporate auxiliary
covariates \citep{cai2019} and heterogeneous measurement errors
\citep{fu2022,gang2026}.}
\textcolor{black}{For the union null considered here, the lfdr must account
for all configurations with at least one nonpositive mean component.
Estimating this probability accurately becomes more difficult as the
number of components increases.}
\textcolor{black}{When these estimates are inaccurate, using them directly to set
rejection thresholds can fail to control FDR.}
\textcolor{black}{This motivates a separate
calibration step that ensures finite-sample FDR control without requiring
accurate lfdr estimates.}

\rev{Symmetry and exchangeability provide a basis for such calibration.}
Knockoff procedures \citep{barber2015,ren2023} and masking
methods \citep{lei2018,leung2022} show how learned rankings can be
combined with finite-sample FDR control.
\rev{\citet{du2023} develop symmetrized data aggregation for dependent
tests, using approximate null symmetry to derive finite-sample FDR
bounds and asymptotic FDR control under regularity conditions.}
\citet{dai2023scale,dai2023split} construct mirror statistics from
split or perturbed data and use their null symmetry to establish
asymptotic FDR control for variable selection. Recent conformal
methods, including PLIS and CLAW \citep{zhao2025claw,zhao2025plis},
pair test scores with calibration scores and preserve their
conditional exchangeability during fitting. These comparisons allow
FDR calibration to depend on a symmetry or exchangeability property
rather than on accurate posterior probability estimates.
\textcolor{black}{Applying these principles to our setting requires mirror
observations that account for correlations among coordinates and support
conservative FDR calibration. The resulting comparisons must remain valid
when the same data are used to learn the ranking.}

\rev{\textcolor{black}{Related work on union null hypotheses includes}
simultaneous knockoffs \citep{dai2023} and the joint mirror procedure
\citep{deng2024}. Their guarantees rely on suitable exchangeability
or mirror conservatism conditions. Establishing such conditions in
our setting requires accounting for two features. First, a null vector can have one nonpositive mean
component and arbitrarily large positive means in the remaining
coordinates. A zero-mean Gaussian reference therefore cannot represent
the whole null. Second, changing the sign of one observed coordinate
generally alters its covariance with the others. The required comparison
must therefore account for the error covariance and remain valid
throughout the union null. In particular, it cannot rely on knowing
which mean component is nonpositive.}

In this paper, we develop an empirical Bayes mirror procedure for
\textcolor{black}{selecting units whose mean vectors have all positive components.}
The procedure learns the distribution of effect vectors across units
and uses estimated posterior null probabilities to rank candidate
discoveries. To calibrate selection, we construct reflected versions
of each observation using its error covariance. The resulting
comparisons classify observations as candidate discoveries or controls,
with a null observation at least as likely to be a control as a
candidate discovery. This property holds without knowing which mean
component is nonpositive. We hide the candidate and control labels
during score estimation to preserve this comparison when learning
from the data. The control count then determines where to cut off
the ranking.

We establish finite-sample FDR control for independent Gaussian
observation vectors with known covariance matrices having nonpositive
off-diagonal entries. The covariance condition covers independent study
estimates and adjacent contrasts formed from independent measurements.
\rev{The guarantee does not require accurate estimation of the posterior
null probabilities used for ranking.} Simulations demonstrate the power gains from learning
the distribution of effects. In the SEDA analysis, the procedure
identifies eight districts with evidence of persistent widening at a
target FDR of 25\%.

The rest of the article is organized as follows.
\rev{Section~\ref{sec:problem-formulation} describes the SEDA data and
formulates the testing problem.} Section~3 develops the procedure
and its theoretical guarantees. Sections~4 and~5 present simulations
and the SEDA analysis, respectively. Section~6 concludes with a
discussion.
% END SECTION 1

% BEGIN SECTION 2
\section{\rev{SEDA data and problem formulation}}
\label{sec:problem-formulation}

\subsection{\rev{Data and scientific question}}
\label{sec:seda-data}

\rev{We analyze the annual mathematics achievement gaps in SEDA 2025.2
\citep{seda2025}, using the administrative-district file that pools
grades 3--8 within each subject and year. The gap is defined as mean
mathematics achievement among non-economically disadvantaged students
minus that among economically disadvantaged students.
The years 2022--2025 refer to school years 2021--22 through 2024--25.
We use the released unshrunk gap estimates and their standard errors,
with economically disadvantaged status defined by each state's classification.
The gaps are on SEDA's cohort-standardized scale, in student
standard-deviation units relative to national reference cohorts
\citep{fahle2026}. Each annual estimate describes the student population
in that district and year, rather than a fixed cohort followed over time.}

\rev{Of the 8,471 districts with a mathematics gap estimate and a positive
standard error in at least one of these years, 5,600 have complete
records for all four years. These districts, located in 33 states,
form our analysis sample. The annual standard errors range from
0.008 to 0.277, indicating substantial variation in measurement
precision. Although 3,415 districts have a larger observed gap in
2025 than in 2022, only 547 have positive observed changes in all
three annual intervals, illustrating the distinction between an
overall increase and persistent widening.}

\rev{Our goal is to identify districts in which the population achievement
gap increased in every annual interval, while controlling the expected
proportion of false discoveries among the selected districts.}

\subsection{\rev{Statistical formulation}}
\label{sec:statistical-formulation}

\rev{To formalize this objective, consider $m$ school districts observed
over $d+1$ consecutive years. For the SEDA sample, $m=5{,}600$ and
the years 2022--2025 provide $d=3$ annual changes.}

For district $i=1,\ldots,m$ and year $t=1,\ldots,d+1$, let
\rev{$g_t^{(i)}$ denote the corresponding population achievement gap.}
Let $\widehat g_t^{(i)}$
and $s_t^{(i)}$ denote its reported estimate and standard error,
respectively.

\rev{To study whether the gap increases in every year, define the true annual
changes and their estimates by
\begin{equation}
 \Delta_j^{(i)}=g_{j+1}^{(i)}-g_j^{(i)},
 \qquad
 X_j^{(i)}=\widehat g_{j+1}^{(i)}-\widehat g_j^{(i)},
 \qquad j=1,\ldots,d.
 \label{eq:pf-changes}
\end{equation}
Write $X^{(i)}=(X_1^{(i)},\ldots,X_d^{(i)})^{\mathsf T}$ and
$\Delta^{(i)}=(\Delta_1^{(i)},\ldots,\Delta_d^{(i)})^{\mathsf T}$.}

\rev{A district has a persistently increasing gap if every coordinate of
$\Delta^{(i)}$ is positive. This leads to the simultaneous tests
\begin{equation}
 H_0^{(i)}:\min_{1\le j\le d}\Delta_j^{(i)}\le0
 \qquad\text{versus}\qquad
 H_1^{(i)}:\Delta_j^{(i)}>0\ \text{for all }j.
 \label{eq:pf-hypotheses}
\end{equation}
A null district has at least one interval in which its population gap
did not increase, regardless of the changes in other intervals. Let
$\mathcal H_0=\{i:\min_j\Delta_j^{(i)}\le0\}$ be the set of null
districts and $\mathcal H_1=\{1,\ldots,m\}\setminus\mathcal H_0$
its complement.}

\rev{To assess these hypotheses, we model the annual estimates as}
\begin{equation}
 \widehat g_t^{(i)}\sim
 \Normal\!\left(g_t^{(i)},\bigl(s_t^{(i)}\bigr)^2\right),
 \qquad i=1,\ldots,m,\quad t=1,\ldots,d+1,
 \label{eq:pf-annual-model}
\end{equation}
independently across districts and years, \textcolor{black}{with fixed unknown
means $g_t^{(i)}$.} \rev{This working model treats the reported standard
errors $s_t^{(i)}>0$ as known error standard deviations and does not
account for uncertainty in their estimation.}

Since $X^{(i)}$ is a linear transformation of the independent Gaussian
estimates in \eqref{eq:pf-annual-model},
\begin{equation}
 X^{(i)}\sim\Normal_d\!\left(\Delta^{(i)},\Sigma^{(i)}\right),
 \qquad i=1,\ldots,m,
 \label{eq:pf-model}
\end{equation}
independently across districts. The covariance matrix has entries
\begin{equation}
 \begin{aligned}
 \Sigma_{jj}^{(i)}
   &=\bigl(s_j^{(i)}\bigr)^2+\bigl(s_{j+1}^{(i)}\bigr)^2,
   &&j=1,\ldots,d,\\
 \Sigma_{j,j+1}^{(i)}=\Sigma_{j+1,j}^{(i)}
   &=-\bigl(s_{j+1}^{(i)}\bigr)^2,
   &&j=1,\ldots,d-1,
 \end{aligned}
 \label{eq:pf-covariance}
\end{equation}
with all other entries equal to zero. Adjacent changes share an annual
estimate with opposite signs, so their errors are negatively correlated
even though the annual errors are independent. The matrices in
\eqref{eq:pf-covariance} are positive definite and satisfy
$\Sigma_{jk}^{(i)}\le0$ for $j\ne k$. Our theory allows any known
covariance matrices with these properties in \eqref{eq:pf-model}.
Each district thus contributes only the pair
$\bigl(X^{(i)},\Sigma^{(i)}\bigr)$.

A selection procedure returns a data-dependent set
$\widehat S\subseteq\{1,\ldots,m\}$ of districts declared to have
persistently increasing gaps. We assess its errors through the false
discovery proportion (FDP) and false discovery rate (FDR):
\begin{equation}
 \FDP(\widehat S)
 =\frac{|\widehat S\cap\mathcal H_0|}{|\widehat S|\vee1},
 \qquad
 \FDR(\widehat S)
 =\E\!\left[\FDP(\widehat S)\right],
 \label{eq:pf-fdr}
\end{equation}
where $|\cdot|$ denotes set cardinality and $a\vee b=\max(a,b)$.
The expectation is over the observation errors and any randomization
used by the procedure. For a prescribed $q\in(0,1)$, our objective is to
\begin{equation}
 \underset{\widehat S}{\operatorname{maximize}}\quad
 \E|\widehat S\cap\mathcal H_1|
 \qquad\text{subject to}\qquad
 \FDR(\widehat S)\le q.
 \label{eq:pf-uniform-fdr}
\end{equation}
We require the FDR bound to hold for every fixed collection of true
change vectors and covariance matrices satisfying the conditions above.
% END SECTION 2

% BEGIN SECTION 3
\section{Covariance-adjusted mirror selection}
\label{sec:mirror-selection}

\textcolor{black}{Using the observed pairs $(X^{(i)},\Sigma^{(i)})$, we construct a
nonnegative score $a^{(i)}$ for each district, with larger scores given
higher priority for selection. Let $\mathcal C=(0,\infty)^d$ denote
the set of vectors with every coordinate positive. For a threshold
$t\ge0$, consider the selection rule}
{\color{black}
\begin{equation}
 \mathcal S(t)=\{i:X^{(i)}\in\mathcal C,\quad a^{(i)}\ge t\}.
 \label{eq:mirror-candidate-threshold}
\end{equation}
}

\textcolor{black}{To choose $t$, we need to estimate the FDP of $\mathcal S(t)$. The
number of selected districts is observed, but the number of false
discoveries among them is unknown. We construct mirrors to identify
observations outside $\mathcal C$ that can serve as controls
\textcolor{black}{for estimating this unknown count}. The aim is to make a null district at least as
likely to contribute a control as a candidate discovery at the same
score threshold.}

\textcolor{black}{Section~\ref{sec:mirror-construction} develops the mirror construction.
Section~\ref{sec:mirror-rule} specifies how the scores may be learned
while preserving this comparison and establishes finite-sample FDR
control for the resulting selection rule.
Section~\ref{sec:mirror-eb} gives an empirical Bayes construction of
the scores.}

\subsection{Mirror construction}
\label{sec:mirror-construction}

For each coordinate $j$, we seek a transformation $R^{(i)[j]}$ such
that the observation and its mirror are pairwise exchangeable for any
fixed true change vector with $\Delta_j^{(i)}=0$:
\begin{equation}
 \bigl(X^{(i)},R^{(i)[j]}X^{(i)}\bigr)
 \stackrel{\mathrm d}{=}
 \bigl(R^{(i)[j]}X^{(i)},X^{(i)}\bigr).
 \label{eq:mirror-exchangeability}
\end{equation}
When $\Delta_j^{(i)}\le0$, we require, for every measurable region
$A\subseteq\mathcal C$,
\begin{equation}
 \Prob\{X^{(i)}\in A\}
 \le \Prob\{R^{(i)[j]}X^{(i)}\in A\}.
 \label{eq:mirror-null-comparison}
\end{equation}
The mirror is then at least as likely as the observation to fall in
any candidate region. Under the null, at least one true change is
nonpositive, but we do not know which. We therefore construct a mirror
for every $j=1,\ldots,d$.

We obtain these properties with the covariance-adjusted reflection
\begin{equation}
 R^{(i)[j]}x
   =x-2\frac{x_j}{\Sigma_{jj}^{(i)}}\Sigma_{\cdot j}^{(i)},
 \qquad x\in\R^d,
 \label{eq:mirror-reflection}
\end{equation}
where $\Sigma_{\cdot j}^{(i)}$ is the $j$th column of
$\Sigma^{(i)}$. The map changes $x_j$ to $-x_j$ and adjusts the other
coordinates according to their error covariances with coordinate $j$.
\rev{When $\Sigma^{(i)}$ is diagonal, this reduces to flipping only
the sign of coordinate $j$.}

\begin{proposition}
\label{prop:mirror-comparison}
Assume the Gaussian observation model \eqref{eq:pf-model}, with a
known, positive-definite covariance matrix $\Sigma^{(i)}$ satisfying
$\Sigma_{jk}^{(i)}\le0$ for $j\ne k$. Then the maps in
\eqref{eq:mirror-reflection} satisfy
\eqref{eq:mirror-exchangeability} and
\eqref{eq:mirror-null-comparison} for every $j=1,\ldots,d$.
Moreover, $R^{(i)[j]}(R^{(i)[j]}x)=x$ for every $x\in\R^d$,
and $R^{(i)[j]}$ maps any $x\in\mathcal C$ to a vector whose
$j$th coordinate is negative and whose remaining coordinates are
positive.
\end{proposition}

To determine which districts can contribute to this comparison, we
examine $X^{(i)}$ and its $d$ reflections
$R^{(i)[1]}X^{(i)},\ldots,R^{(i)[d]}X^{(i)}$ for every district.

{\color{black}
\begin{definition}
\label{def:mirror-eligible-set}
A district is \emph{eligible} if its observed vector or at least one
of its $d$ reflections lies in $\mathcal C$. We define the
\emph{eligible set} by
\begin{equation}
 \mathcal I
 =\{i:X^{(i)}\in\mathcal C\}
 \;\cup\;\bigcup_{j=1}^d
 \{i:R^{(i)[j]}X^{(i)}\in\mathcal C\}.
 \label{eq:mirror-eligible-set}
\end{equation}
Districts in $\mathcal I$ enter the candidate--control comparison.
All remaining districts receive score zero and are excluded from
selection.
\end{definition}
}

For an eligible district, Proposition~\ref{prop:mirror-comparison}
implies that exactly one of these vectors lies in $\mathcal C$;
denote it by $x^{(i)[0]}$. This vector equals $X^{(i)}$ when the
observation has all positive coordinates. Otherwise, it equals
$R^{(i)[j]}X^{(i)}$, where $j$ is the unique negative coordinate of
$X^{(i)}$. We form a list consisting of $x^{(i)[0]}$ and its
$d$ reflections:
\begin{equation}
 x^{(i)[0]},\ x^{(i)[1]},\ldots,x^{(i)[d]},
 \qquad
 x^{(i)[j]}=R^{(i)[j]}x^{(i)[0]},\quad j=1,\ldots,d.
 \label{eq:mirror-list}
\end{equation}
Let $\kappa^{(i)}\in\{0,\ldots,d\}$ identify the actual observation
in this list, so that $X^{(i)}=x^{(i)[\kappa^{(i)}]}$. If
$\kappa^{(i)}=0$, the observed vector lies in $\mathcal C$ and the
district is a candidate discovery in
\eqref{eq:mirror-candidate-threshold}. If $\kappa^{(i)}>0$, the
district serves as a control. Thus each eligible district contributes
one candidate discovery or one control, according to its actual
observation.

Taking any member of \eqref{eq:mirror-list} as the observation
reconstructs the same list. We assign a common score $a^{(i)}$ to
this list, so a candidate discovery and its possible controls are
compared at the same threshold.

\subsection{The selection rule}
\label{sec:mirror-rule}

\rev{To preserve the candidate--control comparison while learning the
scores $a^{(i)}$, we withhold the actual-member labels}
$\kappa^{(i)}$, following the
masking principle used in \citet{lei2018} and \citet{ren2023}.
The information retained for score construction consists of all covariance
matrices, \rev{the eligible set $\mathcal I$ defined in
\eqref{eq:mirror-eligible-set}}, the candidate lists for eligible districts, and the
original observations for ineligible districts. Let $\mathcal M$
denote the sigma-field generated by these data.

We require every score $a^{(i)}$ to be finite and
$\mathcal M$-measurable. All fitting and tuning decisions must therefore
use only the information in $\mathcal M$. In particular, changing which member of any
eligible list was observed, while keeping the lists fixed, must leave
\rev{every score unchanged, including those of other districts.} Section~\ref{sec:mirror-eb} gives a specific
choice of score satisfying this requirement.

After fixing the scores, reveal the labels $\kappa^{(i)}$. At a score
threshold \textcolor{black}{$t\ge0$}, let $N_+(t)$ and $N_-(t)$ count the candidate
discoveries and controls, respectively:
\begin{equation}
 \begin{aligned}
 N_+(t)&=\bigl|\{i\in\mathcal I:\kappa^{(i)}=0,\ a^{(i)}\ge t\}\bigr|,\\
 N_-(t)&=\bigl|\{i\in\mathcal I:\kappa^{(i)}>0,\ a^{(i)}\ge t\}\bigr|.
 \end{aligned}
 \label{eq:mirror-counts}
\end{equation}
Here $N_+(t)=|\mathcal S(t)|$. The probability comparison in
\eqref{eq:mirror-null-comparison} motivates using $N_-(t)$ to estimate
\textcolor{black}{the number of} false discoveries among these selected districts. We use
\begin{equation}
 \widehat{\FDP}(t)=\frac{1+N_-(t)}{N_+(t)\vee1},
 \label{eq:mirror-fdp}
\end{equation}
where the offset $1$ is the finite-sample correction used by the
knockoff+ rule \citep{barber2015}. Let
\textcolor{black}{$\mathcal T=\{a^{(i)}:i\in\mathcal I\}$}.
For the target level $q$, define
\begin{equation}
 T=\min\{t\in\mathcal T:\widehat{\FDP}(t)\le q\},
 \qquad
 \widehat S=\mathcal S(T).
 \label{eq:mirror-selection}
\end{equation}
If no threshold satisfies the condition, set $T=\infty$ and
$\widehat S=\varnothing$. Tied scores are processed together. The next theorem establishes finite-sample FDR control.

\begin{theorem}
\label{thm:mirror-fdr}
Assume the independent Gaussian model \eqref{eq:pf-model}, with known,
positive-definite covariance matrices satisfying
$\Sigma_{jk}^{(i)}\le0$ for $j\ne k$. \textcolor{black}{For any finite,
nonnegative, $\mathcal M$-measurable scores and any
$\mathcal M$-measurable target level $q\in(0,1)$, the selection rule
\eqref{eq:mirror-selection} satisfies}
{\color{black}
\begin{equation}
 \E\{\FDP(\widehat S)\mid\mathcal M\}\le q
 \label{eq:mirror-fdr-guarantee}
\end{equation}
}
for every choice of the true change vectors
$\Delta^{(1)},\ldots,\Delta^{(m)}$.
\textcolor{black}{In particular,  $\FDR(\widehat S)\le q$ when $q$ is fixed.}
\end{theorem}

The theorem permits scores to be learned jointly from all districts through
$\mathcal M$. \rev{Under the stated observation model, their quality can affect power,
while the FDR guarantee requires only the masking condition on how
they are learned.}
\rev{Algorithm~\ref{alg:mirror-selection} summarizes selection given
the scores $a^{(i)}$.}

\begin{algorithm}[H]
\caption{Covariance-adjusted mirror selection}
\label{alg:mirror-selection}
\begin{algorithmic}[1]
 \Require \rev{Observed pairs $\{(X^{(i)},\Sigma^{(i)})\}_{i=1}^m$,
 finite, nonnegative, $\mathcal M$-measurable scores
 $\{a^{(i)}\}_{i=1}^m$, and target FDR level $q$.}
 \Ensure Selected set $\widehat S$.
 \State Construct \rev{the eligible set $\mathcal I$ in
 \eqref{eq:mirror-eligible-set}}, candidate lists, and
 labels $\kappa^{(i)}$ as in Section~\ref{sec:mirror-construction}.
 \State \textcolor{black}{At each distinct score
 $t\in\{a^{(i)}:i\in\mathcal I\}$, compute $N_+(t)$, $N_-(t)$, and
 $\widehat{\FDP}(t)$ using \eqref{eq:mirror-counts} and
 \eqref{eq:mirror-fdp}.}
 \State Choose the threshold $T$ and return $\widehat S$ as in
 \eqref{eq:mirror-selection}; return $\varnothing$ if no threshold
 satisfies the FDP condition.
\end{algorithmic}
\end{algorithm}

\subsection{An empirical Bayes ranking score}
\label{sec:mirror-eb}

\rev{We now use the distribution of changes across districts to construct
an empirical Bayes score based on the posterior probability of
persistent widening. To motivate this score, consider the following
hierarchical working model}
\begin{equation}
 \Delta^{(i)}\overset{\mathrm{iid}}{\sim}G,
 \qquad
 X^{(i)}\mid\Delta^{(i)}
   \sim\Normal_d\!\left(\Delta^{(i)},\Sigma^{(i)}\right),
 \qquad i=1,\ldots,m,
 \label{eq:score-mixture-model}
\end{equation}
independently across districts, where $G$ is a probability distribution
on $\R^d$. This induces a two-group model according to whether
$\Delta^{(i)}\in\mathcal C$.
\rev{This working model motivates the ranking; the FDR guarantee in
Theorem~\ref{thm:mirror-fdr} does not require the effect vectors to
follow a common distribution $G$.}

The relevant ranking quantity is the posterior probability of the union
null. Let $\phi_\Sigma$ denote the centered Gaussian density with
covariance $\Sigma$. The marginal density and lfdr for district $i$
are
\begin{equation}
 \begin{aligned}
 f_G^{(i)}(x)
   &=\int_{\R^d}\phi_{\Sigma^{(i)}}(x-\theta)\,dG(\theta),\\
 \lfdr_G^{(i)}(x)
   &=\Prob\{\Delta^{(i)}\notin\mathcal C\mid X^{(i)}=x\}
     =\frac{\displaystyle\int_{\mathcal C^c}
          \phi_{\Sigma^{(i)}}(x-\theta)\,dG(\theta)}
          {f_G^{(i)}(x)}.
 \end{aligned}
 \label{eq:score-local-fdr}
\end{equation}
Independence across districts implies that
$\lfdr_G^{(i)}(X^{(i)})$ is also the posterior null probability
conditional on all observed vectors.

\rev{To justify giving priority to districts with smaller lfdr values,
consider the oracle setting in which $G$ is known. In this comparison,
both the FDR and the expected number of true discoveries are evaluated under
\eqref{eq:score-mixture-model}. Expectations are taken over both the
random effect vectors and their observations, with the covariance
matrices held fixed.}
The next theorem specializes the ordering argument in
\citet[Theorem~2.1]{heller2021} to our vector observations.

\begin{theorem}
\label{thm:oracle-ranking}
Under \eqref{eq:score-mixture-model}, suppose that $G$ is known.
For any selection rule $\widehat S$ based on the observations, let
$\widehat S_{\mathrm{lfdr}}$ contain the $|\widehat S|$ districts
with the smallest values of $\lfdr_G^{(i)}(X^{(i)})$, breaking ties
by district index. Then
\begin{equation}
 \FDR(\widehat S_{\mathrm{lfdr}})\le\FDR(\widehat S),
 \qquad
 \E|\widehat S_{\mathrm{lfdr}}\cap\mathcal H_1|
 \ge\E|\widehat S\cap\mathcal H_1|.
 \label{eq:oracle-ranking-comparison}
\end{equation}
\end{theorem}

Thus an optimal rule under this model can be chosen to threshold the
lfdr values, with the number of selected districts depending on all
observations. This motivates scores that approximate the lfdr ordering
while satisfying the masking requirement in
Section~\ref{sec:mirror-rule}.

\rev{To estimate the lfdr, we fit $G$ by maximum likelihood using
only the information in $\mathcal M$, following the multivariate
empirical Bayes approach of \citet{soloff2025}.} \rev{Each eligible district contributes one observation, whose label
is hidden. Its likelihood contribution therefore sums the densities
of the $d+1$ possible observations in its list. Since the reflection
maps have absolute Jacobian one, the masked log-likelihood is}
\begin{equation}
 \ell_{\mathrm{mask}}(G)
   =\sum_{i\in\mathcal I}
       \log\!\left\{\sum_{r=0}^d f_G^{(i)}(x^{(i)[r]})\right\}
    +\sum_{i\notin\mathcal I}\log f_G^{(i)}(X^{(i)}).
 \label{eq:mirror-masked-likelihood}
\end{equation}
We approximate $G$ by a mixture of point masses on a finite grid,
as in \citet{jiang2009} and \citet{gangbanerjee2025}. Let $b^{[1]},\ldots,b^{[K]}\in\R^d$ be grid
points chosen using only $\mathcal M$. For weights
$w=(w_1,\ldots,w_K)$, write
\begin{equation}
 G_w=\sum_{k=1}^K w_k\delta_{b^{[k]}},
 \qquad w_k\ge0,\quad \sum_{k=1}^K w_k=1,
 \label{eq:mirror-grid-mixture}
\end{equation}
where $\delta_b$ denotes a point mass at $b$. With the grid fixed,
estimating $G$ reduces to estimating these weights. Substituting
\eqref{eq:mirror-grid-mixture} into
\eqref{eq:mirror-masked-likelihood} gives
\begin{equation}
 \begin{aligned}
 \ell_{\mathrm{mask}}(G_w)
 &=\sum_{i\in\mathcal I}
   \log\!\left\{\sum_{k=1}^K w_k
       \sum_{r=0}^d
       \phi_{\Sigma^{(i)}}(x^{(i)[r]}-b^{[k]})\right\}\\
 &\quad+\sum_{i\notin\mathcal I}
   \log\!\left\{\sum_{k=1}^K w_k
       \phi_{\Sigma^{(i)}}(X^{(i)}-b^{[k]})\right\}.
 \end{aligned}
 \label{eq:mirror-grid-likelihood}
\end{equation}
We estimate the weights by maximizing this log-likelihood subject to
the constraints in \eqref{eq:mirror-grid-mixture}:
\begin{equation}
 \widehat w\in
 \operatorname*{arg\,max}_{\substack{w_k\ge0,\ k=1,\ldots,K\\
                                    \sum_{k=1}^K w_k=1}}
       \ell_{\mathrm{mask}}(G_w),
 \qquad
 \widehat G=\sum_{k=1}^K\widehat w_k\delta_{b^{[k]}}.
 \label{eq:mirror-grid-fit}
\end{equation}
The objective is concave in $w$, and both the grid and the fitted
weights use only the information in $\mathcal M$.

The estimated mixing distribution yields the fitted lfdr
\begin{equation}
 \widehat{\lfdr}^{(i)}(x)
 =\lfdr_{\widehat G}^{(i)}(x)
 =\frac{\displaystyle\sum_{k:\,b^{[k]}\notin\mathcal C}
          \widehat w_k\phi_{\Sigma^{(i)}}(x-b^{[k]})}
        {\displaystyle\sum_{k=1}^K
          \widehat w_k\phi_{\Sigma^{(i)}}(x-b^{[k]})}.
 \label{eq:mirror-fitted-lfdr}
\end{equation}
We then assign each eligible district the score
\begin{equation}
 a^{(i)}=1-\widehat{\lfdr}^{(i)}(x^{(i)[0]}).
 \label{eq:mirror-eb-score}
\end{equation}
Larger scores correspond to smaller estimated lfdr values at the all-positive
vectors $x^{(i)[0]}$. A list receives the same score whichever member
was actually observed. Since $\widehat G$ is fitted using only
$\mathcal M$, the scores are $\mathcal M$-measurable and lie in
$[0,1]$.

The next theorem shows that the fitted lfdr converges to its oracle
counterpart as the grid becomes dense.

\begin{theorem}
\label{thm:masked-consistency}
Assume \eqref{eq:score-mixture-model} with fixed $d\ge2$ and
$\Sigma_{jk}^{(i)}\le0$ for $j\ne k$. Suppose that $G$ is supported
on a fixed compact set $B\subset\R^d$, with
$G(\partial\mathcal C)=0$, where $\partial\mathcal C$ is the boundary
of the positive orthant. Assume that, for constants
$0<\lambda_-\le\lambda_+<\infty$,
$\lambda_- I_d\preceq\Sigma^{(i)}\preceq\lambda_+ I_d$ for all $i$.
Let $\mathcal B_m=\{b^{[1]},\ldots,b^{[K_m]}\}\subset B$ be grids
chosen using only $\mathcal M$, and suppose that
\begin{equation}
 \rho_m=\sup_{\theta\in B}\min_{b\in\mathcal B_m}\|\theta-b\|
 \xrightarrow{p}0.
 \label{eq:mirror-grid-mesh}
\end{equation}
Let $\widehat G_m$ be the estimate in \eqref{eq:mirror-grid-fit}
computed on $\mathcal B_m$. Then, for every $\epsilon>0$, its fitted
lfdr satisfies
\begin{equation}
 \max_{1\le i\le m}\Prob\!\left\{
   \left|\widehat{\lfdr}^{(i)}(X^{(i)})
         -\lfdr_G^{(i)}(X^{(i)})\right|>\epsilon\right\}
 \longrightarrow0\qquad(m\to\infty).
 \label{eq:masked-lfdr-consistency}
\end{equation}
\end{theorem}

%\rev{Under these additional assumptions, the estimated scores converge
%to their oracle counterparts. The grid used in the computations below
%has a fixed resolution, so this asymptotic result does not directly
%apply to that implementation. The finite-sample FDR guarantee in
%Theorem~\ref{thm:mirror-fdr} does not require score consistency.}
% END SECTION 3

% BEGIN SECTION 4
\section{Simulation studies}
\label{sec:simulations}

We compare the FDR and power of five methods under varying signal
strengths and measurement precision.

\subsection{Comparison methods and implementation}
\label{sec:simulation-methods}

All five methods test the hypotheses in \eqref{eq:pf-hypotheses} and
receive the same observations and known covariance matrices. \rev{The target
FDR levels are $q=0.10$ and $0.25$.}
\rev{The first three methods use the same candidate lists and mirror
calibration, differing in their ranking scores. BH and full-data NPMLE
provide alternative approaches to selection.}

\rev{\begin{enumerate}
\item \textbf{EB mirror (proposed method).}
This method combines the empirical Bayes score in
\eqref{eq:mirror-eb-score}, estimated from the masked data, with the
mirror selection rule in Algorithm~\ref{alg:mirror-selection}.

\item \textbf{Fixed mirror.}
This method replaces the learned score with the smallest standardized
coordinate of the all-positive representative:
\begin{equation}
 a_{\mathrm{fix}}^{(i)}
   =\min_{1\le j\le d}
      \frac{x_j^{(i)[0]}}{\sqrt{\Sigma_{jj}^{(i)}}},
 \qquad i\in\mathcal I,
 \label{eq:simulation-fixed-score}
\end{equation}
The score is zero for $i\notin\mathcal I$. Its comparison with EB
mirror assesses the benefit of learning the ranking while keeping
the calibration rule fixed.

\item \textbf{Known-$G$ mirror.}
This reference replaces $\widehat G$ in \eqref{eq:mirror-eb-score}
by the specified mixing distribution $G$ and retains the same mirror
selection rule. Its comparison with EB mirror assesses the effect of
estimating $G$.

\item \textbf{BH.}
For each hypothesis, we calculate the conjunction $p$-value
\begin{equation}
 p^{(i)}=\max_{1\le j\le d}
   \left\{1-\Phi\!\left(
        \frac{X_j^{(i)}}{\sqrt{\Sigma_{jj}^{(i)}}}\right)\right\},
 \label{eq:simulation-conjunction-p}
\end{equation}
where $\Phi$ is the standard normal distribution function. We then
apply the Benjamini--Hochberg procedure \citep{benjamini1995} at
level $q$. This provides a benchmark based on the smallest standardized
coordinate.

\item \textbf{Full-data NPMLE.}
This method fits $G$ without masking and calibrates selection directly
with the estimated lfdr values. Let $\widetilde G$ maximize the ordinary
log-likelihood $\sum_{i=1}^m\log f_G^{(i)}(X^{(i)})$ over weights
on the same support used by EB mirror. This follows the Gaussian
mixture estimation approach of \citet{soloff2025}. Write
$\widetilde L^{(i)}=\lfdr_{\widetilde G}^{(i)}(X^{(i)})$ for the
fitted null probability for hypothesis $i$. We apply the lfdr
step-up rule of \citet[eq.~(10)]{sun2007}: after sorting
$\widetilde L_{(1)}\le\cdots\le\widetilde L_{(m)}$, set
\begin{equation}
 k^*=\max\left\{k:
      \frac1k\sum_{r=1}^k\widetilde L_{(r)}\le q\right\},
 \label{eq:simulation-plugin-rule}
\end{equation}
and reject the $k^*$ hypotheses with the smallest fitted null
probabilities, selecting none if the set is empty.
\end{enumerate}}

\rev{\textit{Shared support and numerical implementation.}
EB mirror and full-data NPMLE use the same support for estimating
$G$.} We collect the $d+1$ vectors in each candidate list for
$i\in\mathcal I$ and the original observations $X^{(i)}$ for
$i\notin\mathcal I$. For each coordinate, we compute the quantiles
of its absolute values in this collection at
$0.10,0.25,0.40,0.55,0.70,0.85,0.95$, and $0.995$.
Each coordinate grid contains 17 values: zero, the eight quantiles,
and their negatives. We take the Cartesian product of the coordinate grids
and add each collected vector having a coordinate outside the
corresponding grid range.

For both methods, we initialize the weights
with 10 EM iterations and then use constrained optimization to
obtain an average log-likelihood within $10^{-4}$ of its maximum.
\rev{For numerical stability, the EB mirror implementation ranks eligible districts by
$-\log\widehat{\lfdr}^{(i)}(x^{(i)[0]})$, capped at a finite value. Before capping, this preserves the ordering
in \eqref{eq:mirror-eb-score} while avoiding rounding scores close
to one; capped ties are processed together.}
Implementation details and numerical checks are provided in the
R code.

\subsection{Varying signal strength}
\label{sec:simulation-signal}

We consider $m=2{,}000$ hypotheses with $d=3$. For signal size
$\mu>0$, the mean vectors $\Delta^{(i)}$ have distribution
\begin{equation}
 G_\mu=0.45\delta_{\mathbf0}
       +0.15\sum_{j=1}^3\delta_{\mu(\mathbf1-e_j)}
       +0.10\delta_{\mu\mathbf1},
 \label{eq:simulation-generating-G}
\end{equation}
where $\mathbf1=(1,1,1)^{\mathsf T}$ and $e_j$ is the $j$th
coordinate vector. Thus, 10\% of the hypotheses are nonnull. Among
the null hypotheses, half have mean zero and half have two positive
coordinates and one zero coordinate.

We randomly permute the indices of these mean vectors, independently
of standard deviations $\sigma^{(i)}$ drawn from
$\operatorname{Unif}(0.5,1.5)$. Conditional on this configuration,
we generate independent observations from \eqref{eq:pf-model} with
\begin{equation}
 \Sigma^{(i)}=(\sigma^{(i)})^2
 \begin{pmatrix}
 1&-1/2&0\\
 -1/2&1&-1/2\\
 0&-1/2&1
 \end{pmatrix}.
 \label{eq:simulation-covariance}
\end{equation}
We vary $\mu$ over $1,1.5,2,2.5$, and $3$.

For each parameter combination, we hold the means and standard
deviations fixed and generate 200 independent sets of measurement
errors. \rev{FDR is estimated by the average realized FDP and power by the
average fraction of the 200 true alternatives selected.}

\begin{figure}[!htbp]
 \centering
 \includegraphics[width=\textwidth]{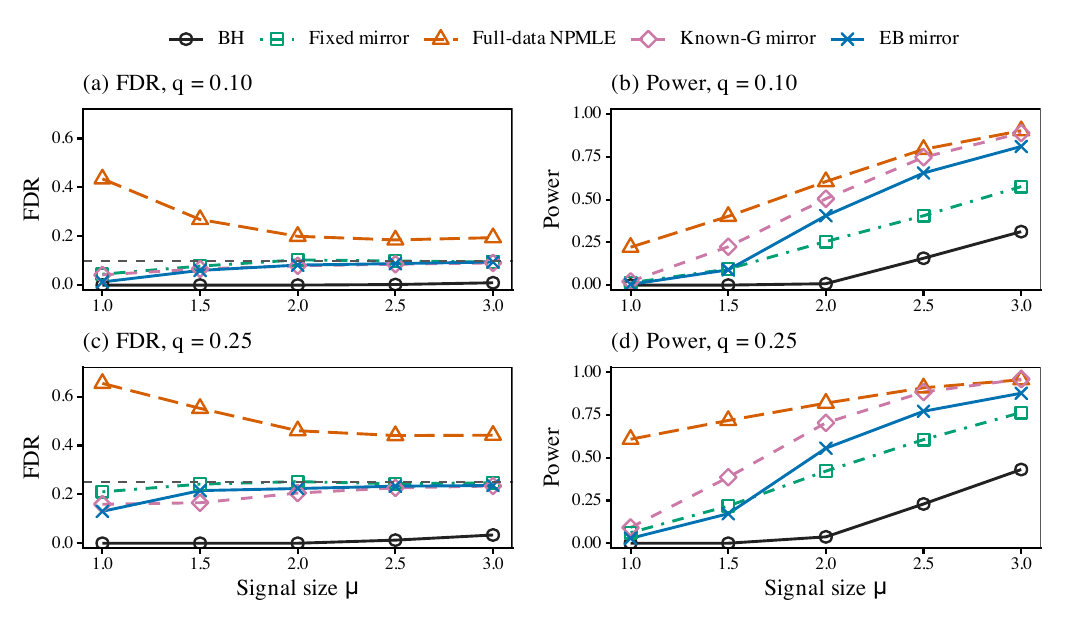}
 \caption{FDR and power as the signal size $\mu$ increases, averaged
 over 200 repetitions with $m=2{,}000$ and $d=3$.
 The upper and lower rows use $q=0.10$ and $q=0.25$, respectively.
 Dashed horizontal lines mark the target FDR levels.}
 \label{fig:simulation-signal}
\end{figure}

\rev{Figure~\ref{fig:simulation-signal} shows empirical FDRs for EB
mirror below both target levels, with the estimates approaching
the targets from below as the signal increases.} The added one in the numerator
of the mirror FDP estimate \eqref{eq:mirror-fdp} makes the calibration
more conservative, particularly when few hypotheses are selected.
Known-$G$ mirror also stays below the targets, while fixed mirror
remains close to them. \rev{Full-data NPMLE has empirical FDR above the target at every
signal size.}

EB mirror is more powerful than fixed mirror for $\mu\ge2$ at both
target levels. At $\mu=2.5$ and $q=0.10$, their powers are $0.655$
and $0.406$, respectively.
 \rev{Both methods use the same mirror calibration, so this comparison
isolates the gain from using the estimated distribution of mean
vectors to rank the hypotheses.} At the two weakest signal sizes, EB mirror and fixed
mirror have comparable power, differing by less than five percentage
points.

BH is the most conservative method and has the lowest power
throughout. Its conjunction $p$-value is calibrated for the least
favorable configuration under the union null and depends only on
the least significant coordinate. Strong evidence in the remaining
coordinates therefore cannot compensate for a weak coordinate,
leading to substantial power loss.

\subsection{Varying measurement precision}
\label{sec:simulation-heterogeneity}

We retain the mean configurations in
\eqref{eq:simulation-generating-G} with $\mu=2.5$ and the covariance
structure in \eqref{eq:simulation-covariance}. To vary measurement
precision, we draw $\sigma^{(i)}$ from
$\operatorname{Unif}(0.5,\sigma_{\max})$, with
$\sigma_{\max}=1,1.25,1.5,1.75$, and $2$. Increasing the upper
endpoint broadens the range of standard errors and increases the
average noise level. At each upper endpoint, we generate 200 new
sets of measurement errors while holding the means and
standard deviations fixed.

\begin{figure}[!htbp]
 \centering
 \includegraphics[width=\textwidth]{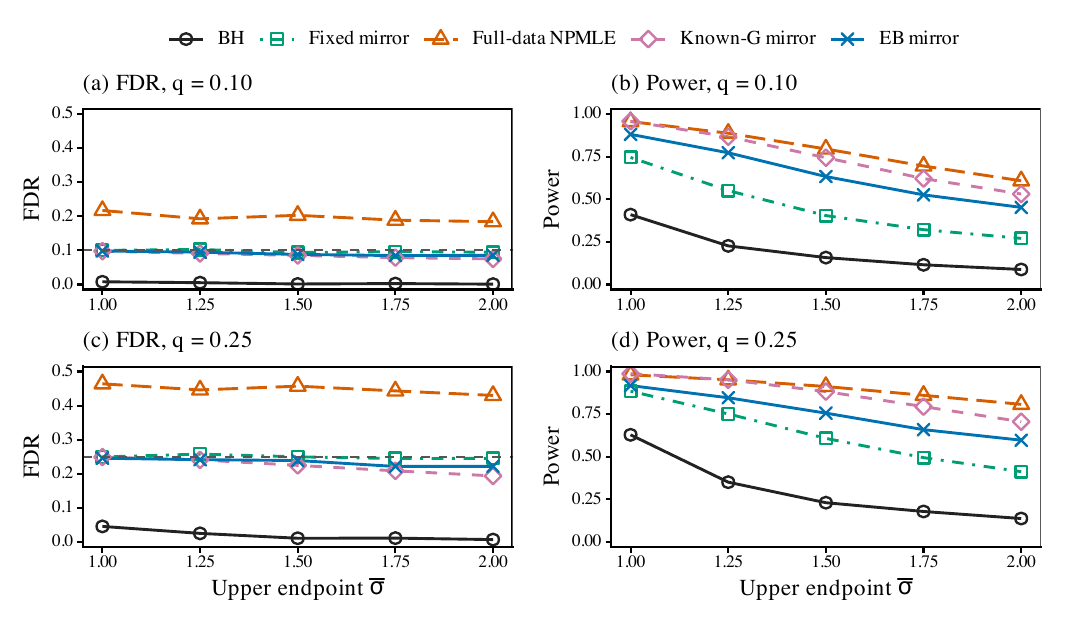}
 \caption{FDR and power as the upper endpoint $\sigma_{\max}$
 of the standard deviation distribution increases, with signal
 size $\mu=2.5$. The remaining settings and plotting conventions
 are the same as in Figure~\ref{fig:simulation-signal}.}
 \label{fig:simulation-heterogeneity}
\end{figure}

The FDR results in Figure~\ref{fig:simulation-heterogeneity} follow
the same pattern as in Section~\ref{sec:simulation-signal}.
Power decreases as measurement errors grow, while EB mirror
remains more powerful than fixed mirror and BH at all five precision
levels.
At $q=0.10$, its power decreases from $0.881$ to $0.452$ as
$\sigma_{\max}$ increases from $1$ to $2$. At the latter setting,
fixed mirror and BH have powers $0.270$ and $0.088$, respectively,
while known-$G$ mirror attains $0.531$.

% END SECTION 4

% BEGIN SECTION 5
\section{Application to the SEDA data}
\label{sec:application}

\rev{We now apply the proposed procedure and comparison methods to the
SEDA sample described in Section~\ref{sec:seda-data}.}

\subsection{\rev{Implementation}}
\label{sec:application-implementation}

\rev{We compare EB mirror, BH, fixed mirror, and full-data NPMLE
at $q=0.10$ and $q=0.25$, using the model in
Section~\ref{sec:statistical-formulation} and the implementations in
Section~\ref{sec:simulation-methods}.}

For EB mirror and full-data NPMLE, we collect the four vectors
\rev{from each of the 1,555 eligible lists and}
the 4,045 original observations without such a list, giving 10,265
vectors. For each coordinate, zero and both signs of the eight
absolute-value quantiles in Section~\ref{sec:simulation-methods}
give 17 values, hence $17^3=4{,}913$ grid points. We add every
distinct collected vector whose absolute value in at least one
coordinate exceeds the corresponding 99.5th percentile. \rev{Adding the 140
vectors satisfying this criterion yields $4{,}913+140=5{,}053$
support points.}

\subsection{Results}
\label{sec:application-results}

\rev{Table~\ref{tab:seda-comparison} summarizes the selection counts.
At $q=0.10$, EB mirror, fixed mirror, and BH select no districts.
At $q=0.25$, EB mirror selects eight districts and fixed mirror
selects four, while BH still selects none. The EB mirror selections
include all four fixed-mirror selections, plus Long Beach Unified,
Washoe County, Wake County Schools, and Cypress-Fairbanks ISD.
Full-data NPMLE yields 109 and 337 selections at $q=0.10$ and
$q=0.25$, respectively. These larger selection counts should be
interpreted in light of the empirical FDR inflation observed for
this method in the simulations in Section~\ref{sec:simulations}.}

\begin{table}[!htbp]
 \caption{\rev{Numbers of districts selected for persistent widening
 among the 5,600 SEDA districts at target FDR levels $q=0.10$ and
 $q=0.25$.}}
 \label{tab:seda-comparison}
 \centering
 \begin{tabular}{lrr}
 \hline
 Method & $q=0.10$ & $q=0.25$ \\
 \hline
 EB mirror & 0 & 8 \\
 Fixed mirror & 0 & 4 \\
 BH & 0 & 0 \\
 Full-data NPMLE & 109 & 337 \\
 \hline
 \end{tabular}
\end{table}

\begin{figure}[!htbp]
 \centering
 \includegraphics[width=0.90\textwidth]{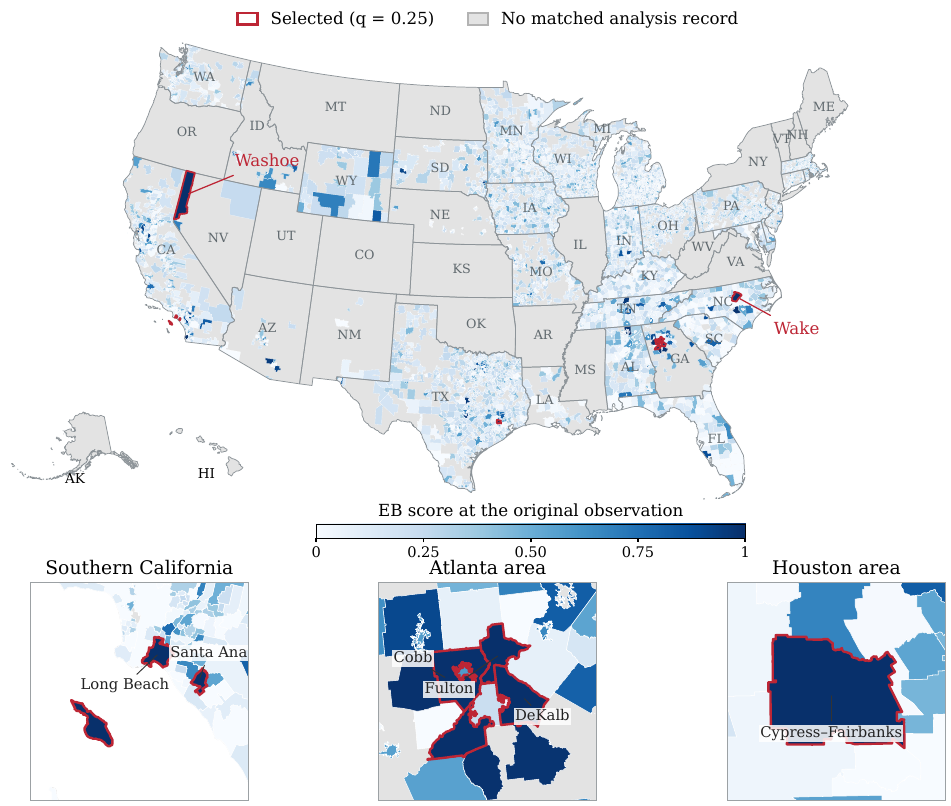}
 \caption{\textcolor{black}{Empirical Bayes support for persistent widening
 and discoveries at $q=0.25$. Blue shading shows
 \mbox{$1-\widehat{\lfdr}^{(i)}(X^{(i)})$}, computed from the masked-data
 estimate $\widehat G$ at each district's original observation.
 All panels use the same $0$--$1$ scale. Red outlines mark the eight
 EB mirror selections; gray areas have no matched analysis record.}
 The map uses 2024 U.S. Census Bureau district boundaries
 \citep{census2024}. Of the 5,600 analyzed districts, 5,286 match a
 boundary identifier, including all eight selected districts.
 \textcolor{black}{Where analyzed elementary and secondary service areas
 overlap, secondary districts are drawn above elementary districts.}}
 \label{fig:seda-map}
\end{figure}

\textcolor{black}{Figure~\ref{fig:seda-map} displays the geographical
distribution of empirical Bayes scores, evaluated at each district's
original observation under the fitted $\widehat G$. Darker shades
indicate greater fitted support for widening in every annual interval\textcolor{black}{.
Red outlines} mark the eight districts selected \textcolor{black}{by} EB mirror at $q=0.25$.
The selected districts
span five states\textcolor{black}{. Three are} in metropolitan Atlanta and two \textcolor{black}{are} in
Southern California.}

Table~\ref{tab:seda-gaps} reports the annual gap estimates for the
selected districts. \rev{In 2022, seven of the eight selected districts
have observed gaps above the median of 0.500 among the 5,600 analyzed
districts. Santa Ana Unified is the only selection below this initial
median. Its gap increases from 0.219 in 2022 to 0.583 in 2025.}
Total observed increases range from 0.151 to
0.505 standard deviations. DeKalb County has the largest increase
in this group, from 1.041 in 2022 to 1.546 in 2025.
Persistent widening is therefore identified across districts
with different initial levels of inequality.

\begin{table}[!htbp]
 \caption{Mathematics gap estimates for EB mirror selections at
 $q=0.25$, in cohort-standardized units. Increase is the change
 \rev{from 2022 to 2025, computed before rounding. Daggers mark fixed-mirror selections.}}
 \label{tab:seda-gaps}
 \centering
 \small
 \setlength{\tabcolsep}{4pt}
 \begin{tabular}{llrrrrr}
 \hline
 District & State & 2022 & 2023 & 2024 & 2025 & Increase \\
 \hline
 Long Beach Unified & CA & 0.589 & 0.709 & 0.714 & 0.801 & 0.212 \\
 Santa Ana Unified$^{\dagger}$ & CA & 0.219 & 0.276 & 0.405 & 0.583 & 0.364 \\
 Cobb County$^{\dagger}$ & GA & 0.752 & 0.969 & 1.041 & 1.082 & 0.330 \\
 DeKalb County$^{\dagger}$ & GA & 1.041 & 1.275 & 1.475 & 1.546 & 0.505 \\
 Fulton County$^{\dagger}$ & GA & 1.153 & 1.274 & 1.345 & 1.393 & 0.239 \\
 Wake County Schools & NC & 1.174 & 1.322 & 1.344 & 1.378 & 0.204 \\
 Washoe County & NV & 0.877 & 0.954 & 1.020 & 1.028 & 0.151 \\
 Cypress-Fairbanks ISD & TX & 0.777 & 0.834 & 0.933 & 0.937 & 0.160 \\
 \hline
 \end{tabular}
\end{table}

\begin{figure}[!htbp]
 \centering
 \includegraphics[width=\textwidth]{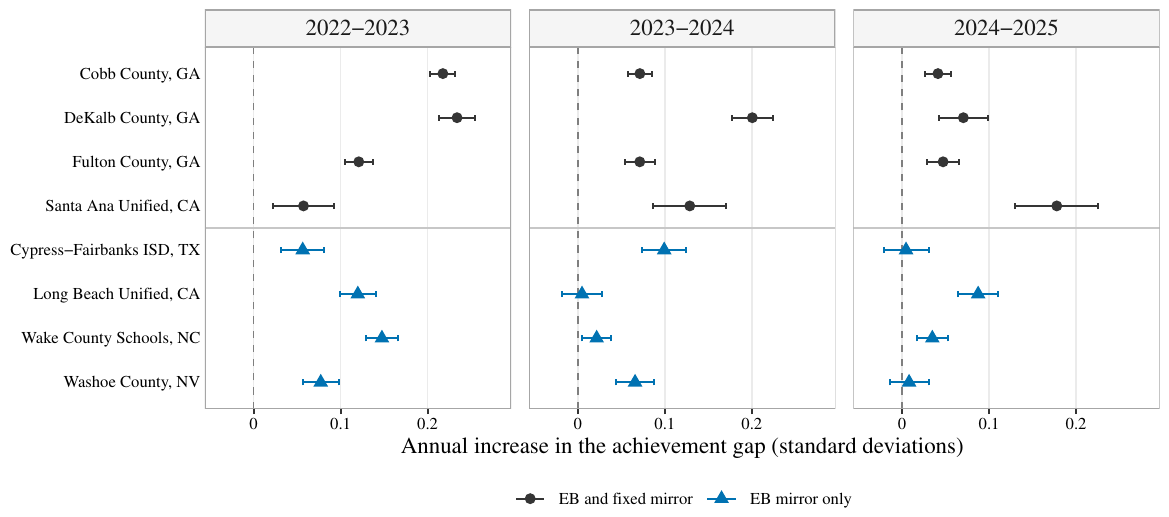}
 \caption{\textcolor{black}{Annual changes in the mathematics achievement
 gap for the eight EB mirror selections at $q=0.25$. Points show adjacent
 annual differences; bars extend one standard error in each direction,
 using the covariance model in Section~\ref{sec:statistical-formulation}.
 Circles mark selections shared with fixed mirror; triangles mark
 selections unique to EB mirror. The bars describe marginal measurement
 uncertainty and are not simultaneous or selection-adjusted confidence
 intervals.}}
 \label{fig:seda-changes}
\end{figure}

\textcolor{black}{Figure~\ref{fig:seda-changes} displays the annual changes
and their standard errors\textcolor{black}{. It distinguishes} the four districts selected
by both mirror methods from the four selected only by EB mirror.
The three Atlanta-area districts have positive but decreasing observed
increments: the gap continues to widen even as the pace slows.
Santa Ana Unified follows a different pattern, with increments of
0.057, 0.129, and 0.178 standard deviations. Its initially smaller gap
and growing increments illustrate why the evolution of a gap provides
information beyond its initial size.}

\textcolor{black}{Each of the four additional EB mirror selections has
at least one interval with an estimated increase smaller than
\textcolor{black}{twice its standard error}. \textcolor{black}{For example, in} Long Beach Unified and Cypress-Fairbanks ISD,
the smallest increases are approximately 0.005, \textcolor{black}{each less
than its standard error}. Their conjunction $p$-values are 0.416 and 0.430, respectively.
Fixed mirror ranks districts by their smallest standardized change.
EB mirror \textcolor{black}{instead} uses the full change vector and the estimated distribution
across districts, which changes the ranking and allows these additional
selections under the same mirror calibration.}

\rev{The selected districts also have relatively precise annual gap
estimates. In each year, all eight districts have reported standard
errors within the lowest decile of the 5,600-district analysis sample.
The annual median standard errors among the selected districts range
from 0.014 to 0.016, compared with 0.092 to 0.096 in the full sample.}
\textcolor{black}{The selections thus reflect both the estimated changes
and their precision\textcolor{black}{. Similar} underlying trajectories may be harder to
detect in districts with larger standard errors.}

% END SECTION 5
\FloatBarrier

% BEGIN SECTION 6
\section{Discussion}
\label{sec:discussion}

We have developed an empirical Bayes mirror procedure for selecting
units with positive effects in every coordinate. The procedure uses
the distribution of effect vectors across units to guide ranking and
adjusts the mirror comparisons for correlated measurement errors.
This allows information to be pooled across units while retaining
finite-sample FDR control under the observation model in
\rev{Section~\ref{sec:statistical-formulation}}, even when the estimated posterior
null probabilities are inaccurate.

The SEDA analysis illustrates the importance of studying how achievement
gaps change over successive years. The selected districts differ
substantially in their initial gap levels, and some have only a small
observed increase in one annual interval. Examining the full sequence
of changes therefore provides information about educational inequality
beyond the size of a gap at a single time point. The resulting selections
identify communities with evidence of sustained divergence between
economically disadvantaged students and their peers during the recovery
period.

Several directions for further research arise from this formulation.
First, the Cartesian grid grows exponentially with the number of annual
changes, making longer follow-up periods computationally demanding.
Models based on a small number of common trajectory patterns may provide
a more efficient way to learn the scores.
The selection rule retains its FDR guarantee when the resulting scores
satisfy the masking requirement in Section~\ref{sec:mirror-rule}.
Second, the covariance matrices determine the mirror transformations
and hence the comparisons used for calibration. Quantifying the effect
of covariance estimation on these comparisons would help extend the
theoretical guarantees to settings with estimated standard errors and
covariances.
\rev{Finally, further applications could assess the procedure's performance
in other conjunction-testing problems. Examples include identifying
effects that are positive across studies and screening for responses
that increase at every dose increment, as discussed in
Section~\ref{sec:introduction}. Extending the calibration to accommodate
dependence across units or non-Gaussian errors would broaden its
applicability in these settings.}

% END SECTION 6

% BEGIN REFERENCES

% END REFERENCES

% BEGIN APPENDIX
\clearpage
\begingroup
\color{black}
\hypersetup{allcolors=black}
\appendix
\section{Proofs}
\label{app:proofs}

\subsection{Proof of Proposition~\ref{prop:mirror-comparison}}
\label{app:mirror-construction}

\begin{proof}
{\color{black}
Fix a district $i$ and coordinate $j$. To simplify notation within
this proof, write $\Sigma=\Sigma^{(i)}$, $\Delta=\Delta^{(i)}$,
$R=R^{(i)[j]}$, and $f(x)=\phi_\Sigma(x-\Delta)$. Let $e_j$ be the
$j$th coordinate vector and set $v=\Sigma_{\cdot j}/\Sigma_{jj}$.
Then $v_j=1$ and $Rx=x-2x_jv$. In particular, $(Rx)_j=-x_j$, so
applying the transformation twice gives
\[
 R(Rx)=x-2x_jv-2(-x_j)v=x.
\]
The matrix of $R$ agrees with the identity matrix in every column
except column $j$. Its $j$th row has $-1$ in column $j$ and zeros
elsewhere. Expanding the determinant along this row leaves an
identity matrix of size $d-1$, and hence $\det R=-1$.
Also, using $\Sigma e_j=\Sigma_{jj}v$, we obtain
\[
 \begin{aligned}
 R\Sigma R^{\mathsf T}
 &=(I_d-2ve_j^{\mathsf T})\Sigma(I_d-2e_jv^{\mathsf T})\\
 &=\Sigma-2\Sigma_{jj}vv^{\mathsf T}
          -2\Sigma_{jj}vv^{\mathsf T}
          +4\Sigma_{jj}vv^{\mathsf T}
 =\Sigma.
 \end{aligned}
\]
Thus the transformation reverses itself, preserves volume, and
preserves the error covariance:
\begin{equation}
 R^2=I_d,\qquad |\det R|=1,\qquad R\Sigma R^{\mathsf T}=\Sigma.
 \label{eq:mirror-identities}
\end{equation}

We next compare the density at $x$ with that at its mirror.
Since $\Sigma^{-1}v=e_j/\Sigma_{jj}$ and
$v^{\mathsf T}\Sigma^{-1}v=1/\Sigma_{jj}$, expanding the two
Gaussian quadratic forms gives
\[
 \begin{aligned}
 &(Rx-\Delta)^{\mathsf T}\Sigma^{-1}(Rx-\Delta)
      -(x-\Delta)^{\mathsf T}\Sigma^{-1}(x-\Delta)\\
 &\quad=-4x_jv^{\mathsf T}\Sigma^{-1}(x-\Delta)
             +4x_j^2v^{\mathsf T}\Sigma^{-1}v\\
 &\quad=-\frac{4x_j(x_j-\Delta_j)}{\Sigma_{jj}}
             +\frac{4x_j^2}{\Sigma_{jj}}
       =\frac{4\Delta_jx_j}{\Sigma_{jj}}.
 \end{aligned}
\]
The normalizing constants of the two densities are identical.
Consequently,
\begin{equation}
 \log\frac{f(x)}{f(Rx)}
   =\frac{2\Delta_jx_j}{\Sigma_{jj}}.
 \label{eq:mirror-density-ratio}
\end{equation}

When $\Delta_j=0$, equation~\eqref{eq:mirror-density-ratio}
gives $f(Rx)=f(x)$ for every $x$. The
change-of-variables formula and $|\det R|=1$ then imply
$RX^{(i)}\stackrel{\mathrm d}{=}X^{(i)}$. Since $R^2=I_d$,
\[
 (RX^{(i)},X^{(i)})
   =(RX^{(i)},R(RX^{(i)}))
   \stackrel{\mathrm d}{=}(X^{(i)},RX^{(i)}),
\]
which proves \eqref{eq:mirror-exchangeability}.
When $\Delta_j\le0$ and $x\in\mathcal C$, the right-hand side of
\eqref{eq:mirror-density-ratio} is nonpositive. Thus
$f(Rx)\ge f(x)$, and, for every measurable $A\subseteq\mathcal C$,
\[
 \Prob\{RX^{(i)}\in A\}
   =\int_A f(Rx)\,dx
   \ge\int_A f(x)\,dx
   =\Prob\{X^{(i)}\in A\}.
\]
This is \eqref{eq:mirror-null-comparison}.

Finally, the nonpositive off-diagonal entries of $\Sigma$ imply
that, for every $x\in\mathcal C$,
\begin{equation}
 (Rx)_j=-x_j<0,\qquad
 (Rx)_k=x_k-2\frac{\Sigma_{kj}}{\Sigma_{jj}}x_j
           \ge x_k>0\quad(k\ne j).
 \label{eq:mirror-signs}
\end{equation}
This proves the remaining assertion of the proposition.
It also explains why an eligible district has a unique all-positive
representative. If $R^{(i)[j]}X^{(i)}\in\mathcal C$, applying the
same reflection again shows that $X^{(i)}$ has its only negative
coordinate at $j$. No other reflection can therefore put it in
$\mathcal C$. Conversely, starting from the $j$th reflected member
of a candidate list and reflecting its unique negative coordinate
recovers $x^{(i)[0]}$. Every member consequently reconstructs the
same list.
}
\end{proof}

\subsection{Proof of Theorem~\ref{thm:mirror-fdr}}
\label{app:mirror-fdr}

\begin{proof}
{\color{black}
We first condition on $\mathcal M$, so that eligibility, the
candidate lists, \textcolor{black}{all scores, and $q$ are fixed}. For an eligible
district, the observation can equal any of the $d+1$ members of
its list. Each map from a member to the all-positive representative
has absolute Jacobian one by \eqref{eq:mirror-identities}.
The conditional probability that member $r$ was observed is
therefore proportional to its density:
\begin{equation}
 \Prob\{\kappa^{(i)}=r\mid\mathcal M\}
   =\frac{f^{(i)}(x^{(i)[r]})}
          {\sum_{s=0}^d f^{(i)}(x^{(i)[s]})},
 \qquad r=0,\ldots,d,
 \label{eq:mirror-label-probabilities}
\end{equation}
where $f^{(i)}(x)=\phi_{\Sigma^{(i)}}(x-\Delta^{(i)})$.
The observations are independent, and the masking is performed
separately for each district. Their joint conditional density
thus factors into these district-specific terms, so the labels
remain independent given $\mathcal M$.

Let $\mathcal I_0=\mathcal I\cap\mathcal H_0$. For every
$i\in\mathcal I_0$, at least one coordinate $j$ satisfies
$\Delta_j^{(i)}\le0$. Equation~\eqref{eq:mirror-density-ratio}
then gives
$f^{(i)}(x^{(i)[j]})\ge f^{(i)}(x^{(i)[0]})$.
The denominator in \eqref{eq:mirror-label-probabilities} is
therefore at least twice its numerator when $r=0$, yielding
\begin{equation}
 \Prob\{\kappa^{(i)}=0\mid\mathcal M\}\le\frac12,
 \qquad i\in\mathcal I_0.
 \label{eq:mirror-null-label-bound}
\end{equation}

It remains to show that this probability bound controls errors
at the data-dependent threshold. We give a direct counting
argument for the knockoff+ calibration
\citep{barber2015,barber2019}. Write
$B^{(i)}=\mathbf1\{\kappa^{(i)}=0\}$, with $B^{(i)}=1$ indicating
a candidate and $B^{(i)}=0$ a control. Let
$V_+(t)=\sum_{i\in\mathcal I_0}B^{(i)}
\mathbf1\{a^{(i)}\ge t\}$ count false discoveries at threshold $t$.
By the definition of $T$,
\begin{equation}
 \FDP(\widehat S)
   \le q\,\frac{V_+(T)}{1+N_-(T)}.
 \label{eq:mirror-fdr-proof}
\end{equation}
If no threshold is selected, both sides are zero. We will show
that the conditional expectation of the ratio on the right is
at most one.

For each $i\in\mathcal I_0$, compute a threshold $T^{(i)}$ by
temporarily setting $B^{(i)}=1$ and leaving all scores and other
labels unchanged. Define
\[
 D^{(i)}
   =1+\sum_{\substack{k\in\mathcal I\\k\ne i}}
         (1-B^{(k)})\mathbf1\{a^{(k)}\ge T^{(i)}\}.
\]
Conditional on $\mathcal M$, neither $T^{(i)}$ nor $D^{(i)}$
depends on $B^{(i)}$. If $B^{(i)}=1$, then $T^{(i)}=T$ and
$D^{(i)}=1+N_-(T)$. Thus, writing
$p^{(i)}=\Prob\{B^{(i)}=1\mid\mathcal M\}$, conditional
independence gives
\[
 \begin{aligned}
 &\E\!\left[
  \frac{B^{(i)}\mathbf1\{a^{(i)}\ge T\}}{1+N_-(T)}
       \,\middle|\,\mathcal M\right]\\
 &\quad=p^{(i)}\E\!\left[
    \frac{\mathbf1\{a^{(i)}\ge T^{(i)}\}}{D^{(i)}}
       \,\middle|\,\mathcal M\right]\\
 &\quad\le\E\!\left[
    \frac{(1-B^{(i)})\mathbf1\{a^{(i)}\ge T^{(i)}\}}{D^{(i)}}
       \,\middle|\,\mathcal M\right],
 \end{aligned}
\]
where the inequality uses $p^{(i)}\le1-p^{(i)}$.

To bound the sum of the last expressions, fix all labels and let
\[
 \mathcal J=\{i\in\mathcal I_0:
          B^{(i)}=0,\ a^{(i)}\ge T^{(i)}\}.
\]
These are null controls that would be selected if their own
labels were changed to candidates. If $\mathcal J$ is nonempty,
all thresholds $T^{(i)}$, $i\in\mathcal J$, are equal.
Indeed, take $i,k\in\mathcal J$ with $T^{(i)}\le T^{(k)}$.
Both scores are at least $T^{(i)}$. At this threshold, changing
either $i$ or $k$ from a control to a candidate produces the same
counts $N_+$ and $N_-$. Hence $T^{(i)}$ also satisfies the FDP
condition when label $k$ is changed. Since $T^{(k)}$ is the
smallest such threshold, $T^{(k)}\le T^{(i)}$, proving equality.

Denote the common threshold by $T_*$. For every $i\in\mathcal J$,
$D^{(i)}=N_-(T_*)$: the added one replaces the contribution of
control $i$. Also, every member of $\mathcal J$ is counted in
$N_-(T_*)$. Therefore,
\[
 \sum_{i\in\mathcal I_0}
   \frac{(1-B^{(i)})\mathbf1\{a^{(i)}\ge T^{(i)}\}}{D^{(i)}}
   =\frac{|\mathcal J|}{N_-(T_*)}\le1.
\]
The sum is zero if $\mathcal J$ is empty. Summing the preceding
conditional expectation inequalities over $i\in\mathcal I_0$
now yields
\begin{equation}
 \E\!\left[\frac{V_+(T)}{1+N_-(T)}
          \,\middle|\,\mathcal M\right]\le1.
 \label{eq:mirror-counting-bound}
\end{equation}
Combining this with \eqref{eq:mirror-fdr-proof} gives
$\E\{\FDP(\widehat S)\mid\mathcal M\}\le q$.

Taking expectations gives
$\FDR(\widehat S)\le\E[q]$, completing the proof.
}
\end{proof}

\subsection{Proof of Theorem~\ref{thm:oracle-ranking}}
\label{app:oracle-ranking}

\begin{proof}
{\color{black}
Let $\mathcal X=(X^{(1)},\ldots,X^{(m)})$ and
$L^{(i)}=\lfdr_G^{(i)}(X^{(i)})$. Independence across districts
implies $\Prob\{i\in\mathcal H_0\mid\mathcal X\}=L^{(i)}$.
Conditional on $\mathcal X$, a selected set $S$ and its size
$r=|S|$ are fixed. Hence
\[
 \E\{\FDP(S)\mid\mathcal X\}
   =\frac{\sum_{i\in S}L^{(i)}}{r\vee1},
 \qquad
 \E\{|S\cap\mathcal H_1|\mid\mathcal X\}
   =r-\sum_{i\in S}L^{(i)}.
\]
Replacing a selected district by an unselected district with a
smaller $L^{(i)}$ decreases the sum in these expressions.
It therefore decreases the conditional expected FDP and
increases the conditional expected number of true discoveries.
Repeating these replacements gives the set of $r$ smallest
lfdr values. Apply this comparison with $r=|\widehat S|$ for
each realization of $\mathcal X$, and then take expectations
to obtain both inequalities in
\eqref{eq:oracle-ranking-comparison}.
}
\end{proof}

\subsection{Proof of Theorem~\ref{thm:masked-consistency}}
\label{app:masked-consistency}

\begin{proof}
{\color{black}
Let $\mathcal P(B)$ be the probability distributions supported on
$B$, and let $\mathcal V$ be the covariance matrices with
eigenvalues in $[\lambda_-,\lambda_+]$ and nonpositive
off-diagonal entries. Both sets are compact, where convergence
of distributions on $B$ means
$\int b\,dH_n\to\int b\,dH$ for every continuous function $b$ on
$B$. In particular, every sequence in $\mathcal P(B)$ has a
subsequence converging in this sense. We first show that the
masked observations determine $G$. We then connect lfdr error
to a loss in expected likelihood and show that a likelihood
maximizer cannot sustain this loss.

\emph{Step 1. Identification from the masked observations.}
For a district with covariance $\Sigma$, let $Z$ record its
eligibility indicator and its retained vector: the original
observation if ineligible, and the all-positive representative
if eligible. The candidate list is determined by $Z$ and
$\Sigma$. For $H\in\mathcal P(B)$, write
\[
 f_{H,\Sigma}(x)=\int_B\phi_\Sigma(x-\theta)\,dH(\theta),
 \qquad
 R_\Sigma^{[j]}x=x-2x_j\Sigma_{\cdot j}/\Sigma_{jj}.
\]
The density $h_{H,\Sigma}$ of $Z$, with respect to counting
measure for the indicator and Lebesgue measure for the vector, is
\[
 \begin{cases}
 f_{H,\Sigma}(x),&\text{if the record is ineligible},\\
 f_{H,\Sigma}(x)+\displaystyle\sum_{j=1}^d
           f_{H,\Sigma}(R_\Sigma^{[j]}x),
       &\text{if the record is eligible}.
 \end{cases}
\]
The sum in the second case follows from the $d+1$ possible
original observations and their unit absolute Jacobians.
These are precisely the contributions to
\eqref{eq:mirror-masked-likelihood}.

Every vector in $Q=(-2,-1)^d$ is ineligible. Indeed, reflecting
coordinate $j$ of an all-negative vector leaves each other
coordinate negative, and there is at least one other coordinate
because $d\ge2$. Thus equality of the masked distributions
implies $f_{H,\Sigma}=f_{G,\Sigma}$ on $Q$.
To see why equality on this open cube is sufficient, write
\[
 f_{H,\Sigma}(x)
   =\frac{e^{-x^{\mathsf T}\Sigma^{-1}x/2}}
          {(2\pi)^{d/2}|\Sigma|^{1/2}}
     \int_B e^{x^{\mathsf T}\Sigma^{-1}\theta
                   -\theta^{\mathsf T}\Sigma^{-1}\theta/2}
             \,dH(\theta).
\]
Since $B$ is bounded, the integral has a power series about
any point in $Q$ that converges for every $x\in\R^d$.
Equality on $Q$ makes all coefficients of the two series equal,
so the densities agree everywhere. Their characteristic
functions therefore satisfy, for every $u\in\R^d$,
\[
 e^{-u^{\mathsf T}\Sigma u/2}
       \int_B e^{\mathrm i u^{\mathsf T}\theta}\,dH(\theta)
 =e^{-u^{\mathsf T}\Sigma u/2}
       \int_B e^{\mathrm i u^{\mathsf T}\theta}\,dG(\theta).
\]
The Gaussian factor is nonzero. Cancelling it shows that $H$
and $G$ have the same characteristic function and hence $H=G$.

\emph{Step 2. Relating lfdr error to expected likelihood.}
Let $\lfdr_{H,\Sigma}(x)$ denote the ratio in
\eqref{eq:score-local-fdr} with mixing distribution $H$ and
covariance $\Sigma$. For a fixed $L<\infty$, define
\[
 D_L(H)=\sup_{\substack{\|x\|\le L\\\Sigma\in\mathcal V}}
          |\lfdr_{H,\Sigma}(x)-\lfdr_{G,\Sigma}(x)|.
\]
We first verify that $H_n\to G$ implies $D_L(H_n)\to0$.
On the compact set of $(x,\Sigma,\theta)$ under consideration,
the Gaussian kernel is continuous and bounded away from zero.
Convergence of its integrals is therefore uniform in
$(x,\Sigma)$: approximate this compact set of $(x,\Sigma)$
values by finitely many points and use uniform continuity
between these points.
For the numerator, the only additional issue is the indicator
$\mathbf1_{\mathcal C^c}(\theta)$. Bound it above and below by
continuous functions taking values in $[0,1]$ that differ only
within distance $\eta$ of $\partial\mathcal C$.
The corresponding integrals converge by the same argument.
Their limiting difference is bounded by a constant times
$G\{\theta:\operatorname{dist}(\theta,\partial\mathcal C)\le\eta\}$,
which tends to zero as $\eta\downarrow0$ because
$G(\partial\mathcal C)=0$. Thus the numerator integrals also
converge uniformly. Dividing by the marginal densities, which
have a uniform positive lower bound, proves the claim.

The expected log-likelihood loss from using $H$ instead of $G$
for a district with covariance $\Sigma$ is
\[
 K_\Sigma(H)
   =\int h_{G,\Sigma}(z)
          \log\frac{h_{G,\Sigma}(z)}{h_{H,\Sigma}(z)}\,dz.
\]
Here integration includes summation over the eligibility
indicator. We need a lower bound for this loss. For any two
densities $p$ and $q$, the inequality $\log u\le u-1$, applied
with $u=\sqrt{q/p}$, gives
\[
 \int p\log(p/q)
   \ge 2\left(1-\int\sqrt{pq}\right)
   =\int(\sqrt p-\sqrt q)^2.
\]
Also, the Cauchy--Schwarz inequality gives
\[
 \left(\int|p-q|\right)^2
 \le \int(\sqrt p-\sqrt q)^2\int(\sqrt p+\sqrt q)^2
 \le4\int(\sqrt p-\sqrt q)^2.
\]
Since observations in $Q$ are retained without transformation,
these two bounds imply
\[
 K_\Sigma(H)\ge
   \frac14\left\{\int_Q
          |f_{H,\Sigma}(x)-f_{G,\Sigma}(x)|\,dx\right\}^2.
\]

We claim that, for every $\epsilon>0$, there is a
$c_{\epsilon,L}>0$ such that
\begin{equation}
 \inf_{\substack{H\in\mathcal P(B):\,D_L(H)\ge\epsilon\\
                  \Sigma\in\mathcal V}}
          K_\Sigma(H)\ge c_{\epsilon,L}.
 \label{eq:masked-kl-separation}
\end{equation}
Otherwise, there would be sequences $H_n,\Sigma_n$ with
$D_L(H_n)\ge\epsilon$ and $K_{\Sigma_n}(H_n)\to0$.
By compactness, pass to a subsequence for which
$H_n\to H_*$ and $\Sigma_n\to\Sigma_*$. The last bound and
continuity of the Gaussian mixtures on the closure of $Q$
give $f_{H_*,\Sigma_*}=f_{G,\Sigma_*}$ on $Q$.
Step 1 then implies $H_*=G$. But the continuity of the lfdr
ratio established above now gives $D_L(H_n)\to0$, a
contradiction. Thus an lfdr error of at least $\epsilon$ entails
a positive expected likelihood loss uniformly over
$\Sigma\in\mathcal V$.

\emph{Step 3. Uniform approximation by the expected likelihood.}
Write $Z^{(i)}$ for the masked record of district $i$, and set
\[
 r_H^{(i)}
   =\log\frac{h_{H,\Sigma^{(i)}}(Z^{(i)})}
                  {h_{G,\Sigma^{(i)}}(Z^{(i)})},
 \qquad
 U_m(H)=\frac1m\sum_{i=1}^m
                  \{r_H^{(i)}-\E r_H^{(i)}\}.
\]
Then $\E r_H^{(i)}=-K_{\Sigma^{(i)}}(H)$. The bounded support
of the means and the eigenvalue bounds imply that, for a
constant $C$ independent of $H$ and $i$,
\[
 |r_H^{(i)}|\le C(1+\|X^{(i)}\|^2).
\]
For example, each Gaussian density with mean in $B$ is bounded
above by a constant and below by
$c_1\exp\{-c_2(1+\|x\|^2)\}$ for fixed positive constants
$c_1,c_2$. The reflection matrices satisfy
$\|R_\Sigma^{[j]}\|\le1+2\lambda_+/\lambda_-$.
Since any list member is obtained from the original observation
by at most two reflections, the same type of bounds holds for
the masked densities.
The Gaussian model also gives uniformly bounded fourth
moments of $\|X^{(i)}\|$.

Fix $L<\infty$ and $\eta>0$. On $\|X^{(i)}\|\le L$, positivity
and uniform continuity of the Gaussian kernels, together with
compactness of $\mathcal P(B)$, give finitely many distributions
$H_1,\ldots,H_N$ such that every $H$ has a representative $H_k$
with $|r_H^{(i)}-r_{H_k}^{(i)}|\le\eta$ for all covariances and
all possible records in this bounded region.
For each of these finitely many $H_k$, independence and the
fourth-moment bound give $\operatorname{Var}\{U_m(H_k)\}\le C'/m$.
Thus $\max_{k\le N}|U_m(H_k)|\xrightarrow{p}0$.
To extend this to all $H$, let
$W_L^{(i)}=(1+\|X^{(i)}\|^2)\mathbf1\{\|X^{(i)}\|>L\}$.
Approximating by the representative on the bounded region
and using the envelope outside it gives
\[
 \sup_H|U_m(H)|
 \le \max_{k\le N}|U_m(H_k)|+2\eta
       +\frac{2C}{m}\sum_{i=1}^m W_L^{(i)}
       +2C\sup_i\E W_L^{(i)}.
\]
The fourth-moment bound implies
$\sup_i\E W_L^{(i)}\to0$ as $L\to\infty$.
The random tail term has the same uniformly vanishing
expectation. Taking $m\to\infty$, then $L\to\infty$ and
$\eta\downarrow0$, proves
\begin{equation}
 \sup_{H\in\mathcal P(B)}|U_m(H)|\xrightarrow{p}0.
 \label{eq:masked-ulln}
\end{equation}

\emph{Step 4. The grid estimate and lfdr convergence.}
Move the mass of any $H\in\mathcal P(B)$ to its nearest grid
points in $\mathcal B_m$, obtaining $H_m$. If $\theta$ is
moved to $b$, then $\|b-\theta\|\le\rho_m$, and direct
expansion gives
\[
 \log\frac{\phi_\Sigma(y-b)}{\phi_\Sigma(y-\theta)}
   =(b-\theta)^{\mathsf T}\Sigma^{-1}(y-\theta)
       -\frac12(b-\theta)^{\mathsf T}\Sigma^{-1}(b-\theta).
\]
Its absolute value is at most $C_1\rho_m(1+\|y\|)$,
uniformly over $\theta,b\in B$ and $\Sigma\in\mathcal V$.
Each vector contributing to the masked likelihood has norm
at most a constant times $\|X^{(i)}\|$. Integrating the
resulting upper and lower density bounds over $H$, and
summing over the possible observations in an eligible list,
therefore yields
\[
 \frac{|\ell_{\mathrm{mask}}(H_m)
               -\ell_{\mathrm{mask}}(H)|}{m}
 \le C_2\rho_m\left(1+\frac1m\sum_{i=1}^m\|X^{(i)}\|\right)
 =o_p(1).
\]
The last equality uses $\rho_m\xrightarrow{p}0$ and the uniformly
bounded first moments of $\|X^{(i)}\|$.
This bound holds for every realized grid and every $H$, so it
also applies when the grid is chosen from $\mathcal M$.
Since $\widehat G_m$ maximizes the likelihood on the grid,
\begin{equation}
 0\le
 \frac{\sup_{H\in\mathcal P(B)}\ell_{\mathrm{mask}}(H)
                   -\ell_{\mathrm{mask}}(\widehat G_m)}{m}
 =o_p(1).
 \label{eq:masked-approximation}
\end{equation}

For fixed $L$ and $\epsilon$, Step 2 gives, whenever
$D_L(H)\ge\epsilon$,
\[
 \begin{aligned}
 \frac{\ell_{\mathrm{mask}}(H)-\ell_{\mathrm{mask}}(G)}{m}
   &=-\frac1m\sum_{i=1}^m K_{\Sigma^{(i)}}(H)+U_m(H)\\
   &\le -c_{\epsilon,L}+\sup_{H'\in\mathcal P(B)}|U_m(H')|.
 \end{aligned}
\]
By Step 3, the last supremum tends to zero in probability.
Such a distribution $H$ consequently has a likelihood below
that of $G$ by at least $mc_{\epsilon,L}/2$ with probability
tending to one. Equation~\eqref{eq:masked-approximation}
prevents $\widehat G_m$ from having this loss. Hence
$D_L(\widehat G_m)\xrightarrow{p}0$ for every finite $L$.

Finally, put $K_B=\sup_{\theta\in B}\|\theta\|$.
The model gives
$\E\|X^{(i)}\|^2\le K_B^2+d\lambda_+$ uniformly in $i$.
For any $\epsilon>0$,
\[
 \begin{aligned}
 &\max_{1\le i\le m}\Prob\!\left\{
    \left|\widehat{\lfdr}^{(i)}(X^{(i)})
           -\lfdr_G^{(i)}(X^{(i)})\right|>\epsilon\right\}\\
 &\qquad\le
   \Prob\{D_L(\widehat G_m)>\epsilon\}
     +\frac{K_B^2+d\lambda_+}{L^2}.
 \end{aligned}
\]
The first term tends to zero for fixed $L$. Letting
$L\to\infty$ proves \eqref{eq:masked-lfdr-consistency}.
}
\end{proof}
\endgroup

% END APPENDIX


\clearpage
\begin{thebibliography}{99}
\bibitem[\protect\citeauthoryear{Barber and Cand\`es}{2015}]{barber2015}
\textsc{Barber, R.~F.} and \textsc{Cand\`es, E.~J.} (2015).
Controlling the false discovery rate via knockoffs.
\emph{Ann. Statist.} \textbf{43} 2055--2085.
\href{https://doi.org/10.1214/15-AOS1337}{doi:\allowbreak{}10.1214/\allowbreak{}15-AOS1337}.

\bibitem[\protect\citeauthoryear{Barber and Cand\`es}{2019}]{barber2019}
\textsc{Barber, R.~F.} and \textsc{Cand\`es, E.~J.} (2019).
A knockoff filter for high-dimensional selective inference.
\emph{Ann. Statist.} \textbf{47} 2504--2537.
\href{https://doi.org/10.1214/18-AOS1755}{doi:\allowbreak{}10.1214/\allowbreak{}18-AOS1755}.

\bibitem[\protect\citeauthoryear{Benjamini and Heller}{2008}]{benjamini2008}
\textsc{Benjamini, Y.} and \textsc{Heller, R.} (2008).
Screening for partial conjunction hypotheses.
\emph{Biometrics} \textbf{64} 1215--1222.
\href{https://doi.org/10.1111/j.1541-0420.2007.00984.x}{doi:\allowbreak{}10.1111/\allowbreak{}j.1541-0420.2007.00984.x}.

\bibitem[\protect\citeauthoryear{Benjamini and Hochberg}{1995}]{benjamini1995}
\textsc{Benjamini, Y.} and \textsc{Hochberg, Y.} (1995).
Controlling the false discovery rate: A practical and powerful approach
to multiple testing.
\emph{J. R. Stat. Soc. Ser. B} \textbf{57} 289--300.
\href{https://doi.org/10.1111/j.2517-6161.1995.tb02031.x}{doi:\allowbreak{}10.1111/\allowbreak{}j.2517-6161.1995.tb02031.x}.

\bibitem[\protect\citeauthoryear{Betth{\"a}user et~al.}{2023}]{betthauser2023}
\textsc{Betth{\"a}user, B.~A.}, \textsc{Bach-Mortensen, A.~M.} and
\textsc{Engzell, P.} (2023).
A systematic review and meta-analysis of the evidence on learning during
the COVID-19 pandemic.
\emph{Nat. Hum. Behav.} \textbf{7} 375--385.
\href{https://doi.org/10.1038/s41562-022-01506-4}{doi:\allowbreak{}10.1038/\allowbreak{}s41562-022-01506-4}.

\bibitem[\protect\citeauthoryear{Cai, Sun and Wang}{2019}]{cai2019}
\textsc{Cai, T.~T.}, \textsc{Sun, W.} and \textsc{Wang, W.} (2019).
Covariate-assisted ranking and screening for large-scale two-sample
inference.
\emph{J. R. Stat. Soc. Ser. B} \textbf{81} 187--234.
\href{https://doi.org/10.1111/rssb.12304}{doi:\allowbreak{}10.1111/\allowbreak{}rssb.12304}.

\bibitem[\protect\citeauthoryear{Chetty, Friedman and Rockoff}{2014}]{chetty2014}
\textsc{Chetty, R.}, \textsc{Friedman, J.~N.} and
\textsc{Rockoff, J.~E.} (2014).
Measuring the impacts of teachers II: Teacher value-added and student
outcomes in adulthood.
\emph{Amer. Econ. Rev.} \textbf{104} 2633--2679.
\href{https://doi.org/10.1257/aer.104.9.2633}{doi:\allowbreak{}10.1257/\allowbreak{}aer.104.9.2633}.

\bibitem[\protect\citeauthoryear{Dai et~al.}{2023a}]{dai2023scale}
\textsc{Dai, C.}, \textsc{Lin, B.}, \textsc{Xing, X.} and
\textsc{Liu, J.~S.} (2023a).
A scale-free approach for false discovery rate control in generalized
linear models.
\emph{J. Amer. Statist. Assoc.} \textbf{118} 1551--1565.
\href{https://doi.org/10.1080/01621459.2023.2165930}{doi:\allowbreak{}10.1080/\allowbreak{}01621459.2023.2165930}.

\bibitem[\protect\citeauthoryear{Dai et~al.}{2023b}]{dai2023split}
\textsc{Dai, C.}, \textsc{Lin, B.}, \textsc{Xing, X.} and
\textsc{Liu, J.~S.} (2023b).
False discovery rate control via data splitting.
\emph{J. Amer. Statist. Assoc.} \textbf{118} 2503--2520.
\href{https://doi.org/10.1080/01621459.2022.2060113}{doi:\allowbreak{}10.1080/\allowbreak{}01621459.2022.2060113}.

\bibitem[\protect\citeauthoryear{Dai and Zheng}{2023}]{dai2023}
\textsc{Dai, R.} and \textsc{Zheng, C.} (2023).
False discovery rate-controlled multiple testing for union null
hypotheses: A knockoff-based approach.
\emph{Biometrics} \textbf{79} 3497--3509.
\href{https://doi.org/10.1111/biom.13848}{doi:\allowbreak{}10.1111/\allowbreak{}biom.13848}.

\bibitem[\protect\citeauthoryear{Deng, He and Zhang}{2024}]{deng2024}
\textsc{Deng, L.}, \textsc{He, K.} and \textsc{Zhang, X.} (2024).
Joint mirror procedure: Controlling false discovery rate for
identifying simultaneous signals.
\emph{Biometrics} \textbf{80} ujae142.
\href{https://doi.org/10.1093/biomtc/ujae142}{doi:\allowbreak{}10.1093/\allowbreak{}biomtc/\allowbreak{}ujae142}.

\bibitem[\protect\citeauthoryear{Dewey et~al.}{2025}]{dewey2025}
\textsc{Dewey, D.~C.}, \textsc{Fahle, E.~M.}, \textsc{Kane, T.~J.},
\textsc{Reardon, S.~F.} and \textsc{Staiger, D.~O.} (2025).
\emph{Pivoting from Pandemic Recovery to Long-Term Reform:
A District-Level Analysis}.
Education Recovery Scorecard, February 2025.
\href{https://cepr.harvard.edu/sites/g/files/omnuum9881/files/cepr/files/pivoting_from_pandemic_recovery_to_long-term_reform_2.9.pdf}{Report}.

\bibitem[\protect\citeauthoryear{Dewey et~al.}{2026}]{dewey2026}
\textsc{Dewey, D.~C.}, \textsc{Fahle, E.}, \textsc{Kane, T.~J.},
\textsc{Reardon, S.} and \textsc{Staiger, D.~O.} (2026).
\emph{From Learning Recession to Learning Recovery: Understanding
the Sources of U.S. K-12 Improvement}.
Education Scorecard, May 2026.
\href{https://educationscorecard.org/wp-content/uploads/2026/05/Education_Scorecard_May_2026_Report.pdf}{Report}.

\bibitem[\protect\citeauthoryear{Du et~al.}{2023}]{du2023}
\rev{\textsc{Du, L.}, \textsc{Guo, X.}, \textsc{Sun, W.} and
\textsc{Zou, C.} (2023).
False discovery rate control under general dependence by symmetrized
data aggregation.
\emph{J. Amer. Statist. Assoc.} \textbf{118} 607--621.
\href{https://doi.org/10.1080/01621459.2021.1945459}{doi:\allowbreak{}10.1080/\allowbreak{}01621459.2021.1945459}.}

\bibitem[\protect\citeauthoryear{Efron et~al.}{2001}]{efron2001}
\textsc{Efron, B.}, \textsc{Tibshirani, R.}, \textsc{Storey, J.~D.} and
\textsc{Tusher, V.} (2001).
Empirical Bayes analysis of a microarray experiment.
\emph{J. Amer. Statist. Assoc.} \textbf{96} 1151--1160.
\href{https://doi.org/10.1198/016214501753382129}{doi:\allowbreak{}10.1198/\allowbreak{}016214501753382129}.

\bibitem[\protect\citeauthoryear{Fahle et~al.}{2026}]{fahle2026}
\textsc{Fahle, E.~M.}, \textsc{Saliba, J.}, \textsc{Min, J.},
\textsc{Reardon, S.~F.}, \textsc{Shear, B.~R.}, \textsc{Shim, J.},
\textsc{Koehler, N.}, \textsc{Kalogrides, D.} and \textsc{Ho, A.~D.} (2026).
\emph{Stanford Education Data Archive Technical Documentation:
SEDA 2025.2}.
Stanford University.
\href{https://purl.stanford.edu/np279jm6134}{SEDA 2025.2 repository}.

\bibitem[\protect\citeauthoryear{Fu et~al.}{2022}]{fu2022}
\textsc{Fu, L.}, \textsc{Gang, B.}, \textsc{James, G.~M.} and
\textsc{Sun, W.} (2022).
Heteroscedasticity-adjusted ranking and thresholding for large-scale
multiple testing.
\emph{J. Amer. Statist. Assoc.} \textbf{117} 1028--1040.
\href{https://doi.org/10.1080/01621459.2020.1840992}{doi:\allowbreak{}10.1080/\allowbreak{}01621459.2020.1840992}.

\bibitem[\protect\citeauthoryear{Gang and Banerjee}{2025}]{gangbanerjee2025}
\textsc{Gang, B.} and \textsc{Banerjee, T.} (2025).
Large-scale multiple testing of composite null hypotheses under
heteroskedasticity.
\emph{Biometrika} \textbf{112} asaf007.
\href{https://doi.org/10.1093/biomet/asaf007}{doi:\allowbreak{}10.1093/\allowbreak{}biomet/\allowbreak{}asaf007}.

\bibitem[\protect\citeauthoryear{Gang et~al.}{2026}]{gang2026}
\textsc{Gang, B.}, \textsc{Fu, L.}, \textsc{James, G.~M.} and
\textsc{Sun, W.} (2026).
Ranking and selection in large-scale inference of heteroscedastic units.
\emph{Ann. Appl. Stat.} \textbf{20} 789--808.
\href{https://doi.org/10.1214/25-AOAS2123}{doi:\allowbreak{}10.1214/\allowbreak{}25-AOAS2123}.

\bibitem[\protect\citeauthoryear{Heller and Rosset}{2021}]{heller2021}
\textsc{Heller, R.} and \textsc{Rosset, S.} (2021).
Optimal control of false discovery criteria in the two-group model.
\emph{J. R. Stat. Soc. Ser. B} \textbf{83} 133--155.
\href{https://doi.org/10.1111/rssb.12403}{doi:\allowbreak{}10.1111/\allowbreak{}rssb.12403}.

\bibitem[\protect\citeauthoryear{Heller and Yekutieli}{2014}]{heller2014}
\textsc{Heller, R.} and \textsc{Yekutieli, D.} (2014).
Replicability analysis for genome-wide association studies.
\emph{Ann. Appl. Stat.} \textbf{8} 481--498.
\href{https://doi.org/10.1214/13-AOAS697}{doi:\allowbreak{}10.1214/\allowbreak{}13-AOAS697}.

\bibitem[\protect\citeauthoryear{Jiang and Zhang}{2009}]{jiang2009}
\textsc{Jiang, W.} and \textsc{Zhang, C.-H.} (2009).
General maximum likelihood empirical Bayes estimation of normal means.
\emph{Ann. Statist.} \textbf{37} 1647--1684.
\href{https://doi.org/10.1214/08-AOS638}{doi:\allowbreak{}10.1214/\allowbreak{}08-AOS638}.

\bibitem[\protect\citeauthoryear{Lei and Fithian}{2018}]{lei2018}
\textsc{Lei, L.} and \textsc{Fithian, W.} (2018).
AdaPT: An interactive procedure for multiple testing with side information.
\emph{J. R. Stat. Soc. Ser. B} \textbf{80} 649--679.
\href{https://doi.org/10.1111/rssb.12274}{doi:\allowbreak{}10.1111/\allowbreak{}rssb.12274}.

\bibitem[\protect\citeauthoryear{Leung and Sun}{2022}]{leung2022}
\textsc{Leung, D.} and \textsc{Sun, W.} (2022).
ZAP: $Z$-value adaptive procedures for false discovery rate control
with side information.
\emph{J. R. Stat. Soc. Ser. B} \textbf{84} 1886--1946.
\href{https://doi.org/10.1111/rssb.12557}{doi:\allowbreak{}10.1111/\allowbreak{}rssb.12557}.

\bibitem[\protect\citeauthoryear{Reardon et~al.}{2026}]{seda2025}
\textsc{Reardon, S.~F.}, \textsc{Fahle, E.~M.}, \textsc{Ho, A.~D.},
\textsc{Shear, B.~R.}, \textsc{Saliba, J.}, \textsc{Min, J.},
\textsc{Shim, J.} and \textsc{Kalogrides, D.} (2026).
\emph{Stanford Education Data Archive, Version SEDA 2025.2}.
\href{https://doi.org/10.25740/np279jm6134}{doi:\allowbreak{}10.25740/\allowbreak{}np279jm6134}.

\bibitem[\protect\citeauthoryear{Ren and Cand\`es}{2023}]{ren2023}
\textsc{Ren, Z.} and \textsc{Cand\`es, E.} (2023).
Knockoffs with side information.
\emph{Ann. Appl. Stat.} \textbf{17} 1152--1174.
\href{https://doi.org/10.1214/22-AOAS1663}{doi:\allowbreak{}10.1214/\allowbreak{}22-AOAS1663}.

\bibitem[\protect\citeauthoryear{Soloff, Guntuboyina and Sen}{2025}]{soloff2025}
\textsc{Soloff, J.~A.}, \textsc{Guntuboyina, A.} and
\textsc{Sen, B.} (2025).
Multivariate, heteroscedastic empirical Bayes via nonparametric
maximum likelihood.
\emph{J. R. Stat. Soc. Ser. B} \textbf{87} 1--32.
\href{https://doi.org/10.1093/jrsssb/qkae040}{doi:\allowbreak{}10.1093/\allowbreak{}jrsssb/\allowbreak{}qkae040}.

\bibitem[\protect\citeauthoryear{Subramanian et~al.}{2017}]{subramanian2017}
\textsc{Subramanian, A.}, \textsc{Narayan, R.},
\textsc{Corsello, S.~M.} et~al. (2017).
A next generation connectivity map: L1000 platform and the first
1,000,000 profiles.
\emph{Cell} \textbf{171} 1437--1452.e17.
\href{https://doi.org/10.1016/j.cell.2017.10.049}{doi:\allowbreak{}10.1016/\allowbreak{}j.cell.2017.10.049}.

\bibitem[\protect\citeauthoryear{Sun and Cai}{2007}]{sun2007}
\textsc{Sun, W.} and \textsc{Cai, T.~T.} (2007).
Oracle and adaptive compound decision rules for false discovery rate
control.
\emph{J. Amer. Statist. Assoc.} \textbf{102} 901--912.
\href{https://doi.org/10.1198/016214507000000545}{doi:\allowbreak{}10.1198/\allowbreak{}016214507000000545}.

\bibitem[\protect\citeauthoryear{U.S. Census Bureau}{2024}]{census2024}
\textsc{U.S. Census Bureau} (2024).
\emph{2024 Cartographic Boundary Files}.
Unified, elementary, and secondary school district boundaries.
\href{https://www.census.gov/geographies/mapping-files/2024/geo/carto-boundary-file.html}{U.S. Census Bureau data repository}.

\bibitem[\protect\citeauthoryear{Zhao and Sun}{2025a}]{zhao2025claw}
\textsc{Zhao, Z.} and \textsc{Sun, W.} (2025a).
A conformalized empirical Bayes method for multiple testing with
side information.
\emph{Preprint}.
\href{https://arxiv.org/abs/2502.19667}{arXiv:2502.19667}.

\bibitem[\protect\citeauthoryear{Zhao and Sun}{2025b}]{zhao2025plis}
\textsc{Zhao, Z.} and \textsc{Sun, W.} (2025b).
False discovery rate control for structured multiple testing:
Asymmetric rules and conformal $Q$-values.
\emph{J. Amer. Statist. Assoc.} \textbf{120} 805--817.
\href{https://doi.org/10.1080/01621459.2024.2359739}{doi:\allowbreak{}10.1080/\allowbreak{}01621459.2024.2359739}.
\end{thebibliography}
\end{document}